\documentclass[a4paper, fleqn, usenatbib, useAMS]{mnras}

\RequirePackage{fixltx2e}

\usepackage{newtxtext,newtxmath}

\usepackage[T1]{fontenc}
\usepackage{ae,aecompl}

\usepackage{graphicx}	
\usepackage{amsmath}	
\usepackage{amssymb}	

\usepackage{CJKutf8}

\newcommand{\iso}[2]{{}^{#2}\mbox{#1}}

\title[MC uncertainties for massive star $s$-process]
{Uncertainties in $s$-process nucleosynthesis in massive stars determined by Monte Carlo variations}
\author[Nishimura~el~al.]
{
N.~Nishimura  \begin{CJK}{UTF8}{ipxm}(西村信哉)\end{CJK}$^{1, 2}$\thanks{e-mail: nobuya.nishimura@yukawa.kyoto-u.ac.jp}\thanks{BRIDGCE UK Network; \url{www.bridgce.ac.uk}},
  R.~Hirschi$^{1, 3}$\footnotemark[2],
  T.~Rauscher$^{4, 5}$\footnotemark[2],
  A.~St.~J.~Murphy$^{6}$\footnotemark[2]\newauthor
  and
  G.~Cescutti$^{5,7}$\footnotemark[2]
\\
  $^1$ Astrophysics Group, Faculty of Natural Sciences, Keele University, Keele ST5 5BG, UK\\
  $^2$ Yukawa Institute for Theoretical Physics, Kyoto University, Kyoto 606-8502, Japan\\
  $^3$ Kavli IPMU (WPI), University of Tokyo, Kashiwa 277-8583, Japan\\
  $^4$ Department of Physics, University of Basel, 4056 Basel, Switzerland\\
  $^5$ Centre for Astrophysics Research, University of Hertfordshire, Hatfield AL10 9AB, UK\\
  $^6$ School of Physics, University of Edinburgh, Edinburgh EH9 3JZ, UK\\
  $^7$ INAF, Osservatorio Astronomico di Trieste, I-34131 Trieste, Italy
}

\date{Accepted 2017 March 17. Received 2017 March 17; in original form 2016 December}

\pubyear{2016}

\begin{document}
\label{firstpage}
\pagerange{\pageref{firstpage}--\pageref{lastpage}}
\maketitle

\begin{abstract}
The $s$-process in massive stars produces the weak component of the $s$-process (nuclei up to $A \sim 90$), in amounts that match solar abundances. For heavier isotopes, such as barium, production through neutron capture is significantly enhanced in very metal-poor stars with fast rotation. However, detailed theoretical predictions for the resulting final $s$-process abundances have important uncertainties caused both by the underlying uncertainties in the nuclear physics (principally neutron capture reaction and $\beta$-decay rates) as well as by the stellar evolution modeling. In this work, we investigated the impact of nuclear-physics uncertainties relevant to the $s$-process in massive stars. Using a Monte-Carlo based approach, we performed extensive nuclear reaction network calculations that include newly evaluated upper and lower limits for the individual temperature dependent reaction rates. We found that most of the uncertainty in the final abundances is caused by uncertainties in the neutron capture rates, while $\beta$-decay rate uncertainties affect  only a few nuclei near $s$-process branchings. The $s$-process in rotating metal-poor stars shows quantitatively different uncertainties and key reactions, although the qualitative characteristics are similar. We confirmed that our results do not significantly change at different metallicities for fast rotating massive stars in the very low metallicity regime. We highlight which of the identified key reactions are realistic candidates for improved measurement by future experiments. 
\end{abstract}




\clearpage


\section{Introduction}
\label{sec-intro}

The slow neutron capture process, the {\it $s$-process} \citep[see, {\it e.g.}][]{1957RvMP...29..547B, 1965ApJS...11..121S}, is one of the major nucleosynthesis processes. It produces heavy elements beyond iron by sequences of neutron captures and $\beta$-decays. The principal characteristic of this process is that the time scale for neutron capture is generally much slower ($\gg 1~\mbox{yr}$) than for $\beta$-decay, even for ground states of nuclei near stability, resulting in the main $s$-process nucleosynthesis path to lie along the ``$\beta$-stable valley'' of the chart of nuclei. The $s$-process begins from seed nuclei, which are mainly the $\iso{Fe}{56}$ nuclei initially present in the star, and proceeds by capturing neutrons released from ($\alpha$,n) reactions on lighter nuclei (lighter that than the iron group) occurring in stellar nuclear burning. Massive stars ($\gtrsim 10~M_{\odot}$) are considered to be the main astronomical site for the {\it weak} $s$-process (hereafter the {\it ws}-process), producing the {\it weak} component of the $s$-process, responsible for nuclides with mass numbers up to $A \sim 90$ \citep[{\it e.g.},][]{1990A&A...234..211P, 2010ApJ...710.1557P}. On the other hand, thermal pulses in low mass asymptotic giant branch stars are the site of the {\it main} $s$-process, producing the {\it main} component \citep[see {\it e.g.},][and references therein]{2011RvMP...83..157K, 2015MNRAS.449..506B}.

Considering the {\it ws}-process, this occurs in helium-core and carbon-shell burning phases of massive stars. The evolution of the star is governed by several fusion reactions, {\it e.g.}, the triple-$\alpha$ reaction \citep{2011ApJ...741...61S, 2012PThPh.127..171K}, $\iso{C}{12} + \alpha \rightarrow \iso{O}{16}$ in helium burning, and $\iso{C}{12} + \iso{C}{12}$ in carbon burning \citep[][]{2012MNRAS.420.3047B, 2013ApJ...762...31P}. The impact of uncertainty for these reactions on the {\it ws}-process has been investigated for stellar temperatures $\sim 200~\mbox{MK} \equiv 17.2~\mbox{keV}$ \citep[for recent studies, see,][and references therein]{2009ApJ...702.1068T, 2015MNRAS.447.3115J}.  The main neutron source reaction for massive star evolution is $\iso{Ne}{22}(\alpha, \mbox{n})\iso{Mg}{25}$, while the competing $\iso{Ne}{22}(\alpha, \gamma)\iso{Mg}{26}$ reaction reduces the production of source neutrons, and, notably, has a reaction rate that is still uncertain. Recent studies \citep{2002ApJ...576..323R,2002JNSTS..2...512P,2012PhRvC..85f5809L,2014AIPC.1594..146N} have revealed that, for a reasonable range of updated nuclear physics properties, the ratio of the rates of the $\iso{Ne}{22}(\alpha, {\mbox n})\iso{Mg}{25}$ reaction to the $\iso{Ne}{22}(\alpha, \gamma)\iso{Mg}{26}$ reaction has a significant impact on the final {\it ws}-process products. 

At solar metallicity, rotation-induced mixing enhances {\it ws}-process production by a factor of up to a few, but the general production and the nucleosynthesis path are the same in rotating as in non-rotating models. It is thus not necessary to consider additional models for rotating stars to study the impact of nuclear uncertainties on the {\it ws}-process at solar metallicity. As the metallicity of stars decreases, however, rotation-induced mixing has stronger effects on nucleosynthesis, including the {\it ws}-process \citep{2008ApJ...687L..95P}. Stellar evolution calculations that include rapid rotation \citep{2012A&A...538L...2F, 2016MNRAS.456.1803F} show strong mixing between the helium-burning core and the hydrogen-burning shell. Firstly, this mixes primary $\iso{C}{12}$ and $\iso{O}{16}$ into the hydrogen-burning shell, leading to the production of additional $\iso{N}{14}$ in this shell via the CNO cycle. Secondly, this $\iso{N}{14}$ is then mixed back into the helium-burning core, at which point it immediately converts via the $\iso{N}{14}(\alpha, \gamma)\iso{F}{18}({\rm e}^+\nu_{\rm e})\iso{O}{18}(\alpha, \gamma)\iso{Ne}{22}$ series of reactions into $\iso{Ne}{22}$, {\it i.e.}, the fuel for the main neutron source reaction. Finally, at the end of He-core burning, $\iso{Ne}{22}(\alpha,\mbox{n})\iso{Mg}{25}$ reactions release large numbers of neutrons. \citep{2012A&A...538L...2F, 2016MNRAS.456.1803F}. Due to a larger neutron exposure, combined with a lower seed abundance, the production of heavier nuclei with mass numbers $A > 100$ is enhanced as compared to the ``standard'' {\it ws}-process that only produces nuclei up to $A \sim 90$.

This {\it enhanced} weak $s$-process (denoted here as the {\it es}-process), which is described as a ``non-standard'' $s$-process in \cite{2012A&A...538L...2F}, has a significant contribution to the chemical evolution of galaxies. Although the {\it es}-process is considered to be active only in very metal-poor stars, it is possibly a source of heavy elements ({\it e.g.} barium) in the early Universe. While early studies have ignored the contributions from massive stars \citep[see,][]{1992ApJ...387..263R}, as pointed out in \cite{2011Natur.472..454C, 2011Natur.474..666C} it has since been shown that the {\it es}-process can have important impact on chemical enrichment in early galaxies. Furthermore, {\it es}-process production coupled with an $r$-process production can explain the dispersion observed in the light neutron capture elements over the heavy neutron capture elements in Galactic halo stars \citep{2013A&A...553A..51C}. 

For the {\it es}-process, in addition to the neutron source reactions, it is important to consider the effect of the $\iso{O}{16}({\mbox n}, \gamma)\iso{O}{17}$ neutron poison reaction. The net efficiency of this poison reaction is determined by the ratio between the reaction paths $\iso{O}{16}({\mbox n}, \gamma)\iso{O}{17}(\alpha, {\mbox n})\iso{Ne}{20}$ and $\iso{O}{16}({\mbox n}, \gamma)\iso{O}{17}(\alpha, \gamma)\iso{Ne}{21}$ \citep[see,][]{2012A&A...538L...2F, 2016MNRAS.456.1803F, 2014AIPC.1594..146N}. This is poorly known because, in particular, the rate of the $\iso{O}{17}(\alpha, \gamma)\iso{Ne}{21}$ reaction is experimentally undetermined, and evaluated reaction rates are different by several orders of magnitude \citep{2010nuco.confE..45T, 2011PhRvC..83e2802B, 2013PhRvC..87d5805B}. Consequently, the final abundances of the {\it es}-process are significantly influenced by this uncertainty as shown in previous studies \citep{2012A&A...538L...2F, 2014AIPC.1594..146N}.

In the present study, we investigate the impact of nuclear-physics uncertainty relevant to the $s$-process  ({\it ws}- and {\it es}-processes) in massive stars. We focus on $(\mbox{n}, \gamma)$ reactions and $\beta$-decay on the path of $s$-process nucleosynthesis. We perform comprehensive rate variations using the {\tt PizBuin} Monte-Carlo (MC) framework coupled with a reaction network code, described previously in \citet{2016MNRAS.463.4153R}. Analysing the results of the MC calculations, we determine the important reactions and decays that are the dominant sources of uncertainty for the production of ``key'' elements. We highlight such reactions that should be investigated in future nuclear physics studies.

Importantly, the uncertainties used for the nuclear reaction rates include a temperature-dependence due to the relative contributions of ground states and excited states. Following the prescription of \cite{2011ApJ...738..143R} and \cite{2012ApJS..201...26R}, we apply temperature dependent uncertainties even for experimentally evaluated $(\mbox{n}, \gamma)$ reaction rates. This leads to a higher uncertainty compared to ground-state contributions alone. We use a similar approach for the $\beta$-decay rates, also including dependency on temperature, based on nuclear partition functions.

The present paper has the following structure.  In Section~\ref{sec-methods}, we describe the stellar evolution models and the nuclear reaction networks used in this study, as well as the method of MC simulation with the evaluation of uncertainty for the reaction rates. The results of standard nucleosynthesis and rate variation with the MC approach for {\it ws}- and {\it es}-processes are shown in Section~\ref{sec-wsproc} and in Section~\ref{sec-esproc}, respectively. We discuss the possibilities for improvement of nuclear data in Section~\ref{sec-nuc-data}. Section~\ref{sec-summary} is devoted to discussion and conclusions.


\section{Nucleosynthesis and Monte-Carlo methods}
\label{sec-methods}

\subsection{Stellar evolution models}
\label{sec-evolution}

Contemporary nucleosynthesis calculations for the {\it ws}-process in massive stars use full stellar evolution models. The complete nucleosynthesis is calculated either fully coupled \citep[see {\it e.g.}][]{2016MNRAS.456.1803F} or using a post-processing approach \citep{2016ApJS..225...24P}. In a Monte Carlo framework, however, the calculations need to be repeated many ({\it e.g.} $10,000$) times and using fully coupled networks in stellar models would be computationally extremely expensive.

To establish a more tractable approach, we have thus created a {\it single-zone trajectory} that mimics the average thermodynamic (density and temperature) history and nucleosynthesis occurring during core helium and carbon shell burning. The trajectory was chosen in such a way that an equal amount of $^{22}$Ne burnt in the trajectory and the full stellar model. This simplification is reasonable because the {\it ws}-process is produced in large convective zones in massive stars, in which quantities vary smoothly and not too significantly. This procedure was used in several studies in the past \citep{2008ApJ...687L..95P, 2008nuco.confE..83H}. The trajectory used in this work was extracted from a solar metallicity $25~M_\odot$ model \citep{2004A&A...425..649H}, and was chosen because it corresponds roughly to the average {\it ws}-process production in massive stars weighted over the initial mass function.

\begin{figure}
  \centering
  \includegraphics[width=\hsize]{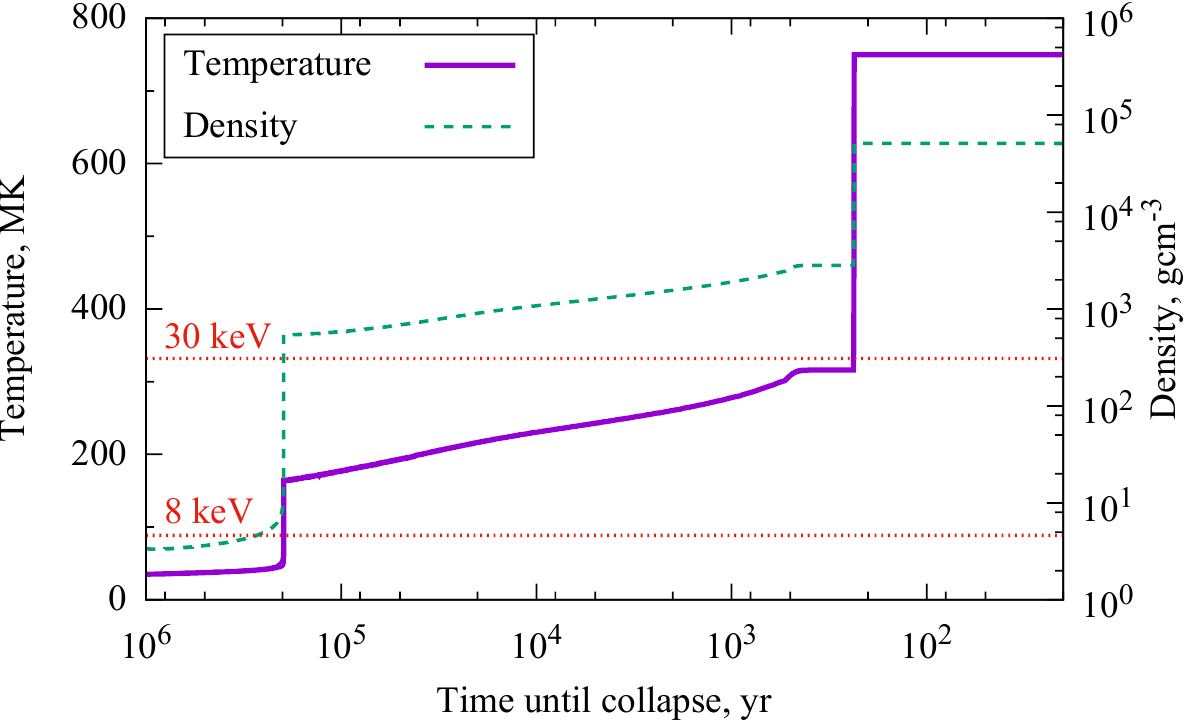}
  \caption{\label{fig-star_hydro}The density and temperature evolution of the single trajectory of a $25 M_\odot$ star model \citep{2004A&A...425..649H}. The temperature in MK and the density in $\mbox{g~cm}^{-3}$ are shown. The temperature region between $8$ and $30$~keV, relevant for the {\it ws}-process, is delimited by horizontal dashed lines.}
\end{figure}

In Figure~\ref{fig-star_hydro}, we show the temporal evolution of the density and temperature for the adopted trajectory. The trajectory follows the core-hydrogen, core-helium and shell-carbon burning phases. It thus covers the entire evolution of the star from the zero-age main-sequence to core collapse (the carbon burning shell is still active at core-collapse). Since the precise choices of initial metallicity and rotation affect the nucleosynthesis yields much more strongly than the stellar structure, we may, without loss of generality of the discussion on the nuclear uncertainty in the {\it ws}-process, assume one representative thermodynamic trajectory, and change rotation and metallicity parameters as required.

Adopting this thermodynamic ({\it i.e.}, the temperature and density) trajectory, we consider a range of initial compositions from very metal-poor stars to solar metallicity. Metallicities are indicated by $Z_{\rm m} = 1 - X_{\rm H} - X_{\rm He}$, with $X_\mathrm{H}$ and $X_\mathrm{He}$ being the mass fractions of hydrogen and helium, respectively. We adopt $Z_{\rm m} = 1.4 \times 10^{-2} \equiv Z_\odot$ as the solar metallicity and considered four additional metallicity models. The adopted values are presented in Table~\ref{tab-param} and are denoted by {\tt z0} ($Z_\odot$), {\tt z1}, {\tt z2}, {\tt z3}, and  {\tt z4}. In addition, the effect of rotation-induced mixing is considered by means of adding extra $\iso{N}{14}$ to the initial composition. This primary $\iso{N}{14}$ immediately converts to $\iso{Ne}{22}$ at the start of core helium-burning and enhances the {\it ws}-process production. This causes the {\it es}-process in the rotating massive stars at low metallicities. Such a simplified approach has been shown to provide consistent results in nucleosynthesis similar to more sophisticated evolution calculations \citep[see,][]{2012A&A...538L...2F, 2016MNRAS.456.1803F}. Following \cite{2012A&A...538L...2F}, we choose a mass fraction of $X(\iso{N}{14}) = 0.01$ for the fastest rotating case, and consider five values for the initial $\iso{N}{14}$ to represent a range of rotation, from non-rotating, {\tt r0}, to a maximum rotation, {\tt r4}, see Table~\ref{tab-param}.

In this study, the stellar models are set by choosing various combinations of metallicity and effective rotation, while always using the same thermodynamic trajectory. In terms of our notation, the stellar model at solar metallicity without rotation is denoted {\tt z0r0}, while the fast rotating metal-poor star is {\tt z2r4}. The {\tt z0r0} model shows typical {\it ws}-process final abundances, whereas rotating metal-poor stars show abundance patterns that differ due to the {\it es}-process. Nucleosynthesis results are presented in Sections~\ref{sec-wsproc} and \ref{sec-esproc}. 

\begin{table}
 \caption{ \label{tab-param}Parameters of metallicity and rotation, defining the stellar models used. The values of {\tt z0} to {\tt z4} denote the initial metallicity $Z_{\rm m}$. The strength of rotation, {\tt r0} to {\tt r4}, is characterized by the mass fraction of primary $\iso{N}{14}$.}
 \begin{tabular}{@{}l|cccccc}
  \hline
   & {\tt z0} & {\tt z1} & {\tt z2} & {\tt z3} & {\tt z4} \\
  \hline
   & $1.4 \times 10^{-2}$ & $6.0 \times 10^{-3}$ & $1.0 \times 10^{-3}$
  & $1.0 \times 10^{-4}$ & $1.0 \times 10^{-5}$ \\
  \hline
  \\
  \\
  \hline
   & {\tt r0} & {\tt r1} & {\tt r2} & {\tt r3} & {\tt r4}\\
  \hline
  & $0$ &  $5 \times 10^{-4}$
  & $1 \times 10^{-3}$ & $5 \times 10^{-3} $ & $1\times 10^{-2}$ \\
  \hline
 \end{tabular}

\medskip
\end{table}

\begin{table}
 \caption{ \label{tab-network}Important reactions related to neutron production and consumption in the {\it ws}- and {\it es}-processes with their references.}
\begin{center}
 \begin{tabular}{@{}cc}
  \hline
   Reaction &  Rate taken from \\
   \hline
   $\iso{Ne}{22}(\alpha, {\mbox n})\iso{Mg}{25}$ &  \cite{2001PhRvL..87t2501J} \\
   $\iso{Ne}{22}(\alpha, \gamma)\iso{Mg}{26}$ &  \cite{1999NuPhA.656....3A} \\
   $\iso{O}{17}(\alpha, {\mbox n})\iso{Ne}{20}$ &  \cite{1999NuPhA.656....3A} \\
   $\iso{O}{17}(\alpha, \gamma)\iso{Ne}{21}$ & \cite{1988ADNDT..40..283C}$\times 0.1$ $^{\mathrm *a}$ \\
  \hline
 \end{tabular}
 \end{center}
 $^{\mathrm *a}$ A modified rate, set at 0.1 of the rate of \cite{1988ADNDT..40..283C}, has been used for the MC calculations (see text for details). 
\end{table}

\subsection{Nuclear reaction networks}
\label{sec-network}

The reaction network consists of $943$ isotopes including all reactions relevant to the $s$-process, {\it e.g.}, fusion reactions of lighter isotopes as well as (n,$\gamma$) reactions and $\beta$-decays or electron captures of heavy nuclei. The numerical values of theoretical and experimental reaction rates are taken from \cite{2000ADNDT..75....1R} and \cite{2010ApJS..189..240C}. The majority of (n,$\gamma$) experimental rates are taken from the KADoNiS compilation \citep{2006AIPC..819..123D}, which provides the standard nuclear reaction rate input. We adopt temperature dependent $\beta$-decay rates from \cite{1987ADNDT..36..375T} and \cite{1999A&A...342..881G} as provided in \cite{2005A&A...441.1195A} and \cite{2013A&A...549A.106X}. Original data of the decay rates are given by numerical tables. These $\beta$-decays rates are dependent on the temperature, and we express them by a seven-parameter fitting formula
\citep[][]{2000ADNDT..75....1R}:
\begin{equation}
	\label{eq-reaclib}
	\begin{split}
	\lambda(T_9) = \exp (a_0 &+ a_1 {T_9}^{-1} + {a_2} {T_9}^{-1/3} \\
	&+ {a_3} {T_9}^{1/3} + {a_4} {T_9} + {a_5} {T_9}^{5/3} + {a_6} \ln{T_9}) \ ,
	\end{split}
\end{equation}
where $a_0$--$a_6$ are constant coefficients and $T_9$ is the temperature in $10^{9}$~K (GK).

Neutron sources and neutron poisons are key reactions in $s$-process nucleosynthesis. We adopt \cite{2001PhRvL..87t2501J} for $\iso{Ne}{22}(\alpha, \mbox{n})\iso{Mg}{25}$ and \cite{1999NuPhA.656....3A} for $\iso{Ne}{22}(\alpha, \gamma)\iso{Mg}{26}$, respectively. In addition to the neutron source reactions, abundant $\iso{O}{16}$ in the helium core and the carbon shell is a strong neutron absorber. Thus, it may be a strong neutron poison. Although the poison reaction, $\iso{O}{16}(\mbox{n},\,\gamma)\iso{O}{17}$, is relatively well determined, rates of ($\alpha$,\,n) and $(\alpha, \gamma)$ on the produced $\iso{O}{17}$ have large uncertainties. The net efficiency of the poison reactions is determined by the competition between the reaction combination of $\iso{O}{16}(\mbox{n}, \gamma)\iso{O}{17}(\alpha, \gamma)\iso{Ne}{21}$ and $\iso{O}{16}(\mbox{n}, \gamma)\iso{O}{17}(\alpha, \mbox{n})\iso{Ne}{20}$. Only the latter sequence makes $\iso{O}{16}$ a neutron poison. Following the previous study \citep{2012A&A...538L...2F}, we adopt the $\iso{O}{17}(\alpha, \mbox{n})\iso{Ne}{20}$ rate from \cite{1999NuPhA.656....3A}, while we use the rate of \cite{1988ADNDT..40..283C} for $\iso{O}{17}(\alpha,\gamma)\iso{Ne}{21}$ divided by a factor of $10$. The choices for which rates to use for the main neutron source and poison reactions are summarized in Table~\ref{tab-network}.

\subsection{Reaction rate variation}
\label{sec-pizbuin}

The Monte-Carlo (MC) method, which treats physical uncertainty through the use of repeated random variation, is a robust methodology to examine nucleosynthesis uncertainties \citep[see, {\it e.g.}][]{2014arXiv1409.5541I, 2016MNRAS.463.4153R}. We use the {\tt PizBuin} MC driver coupled with a nuclear reaction network. This framework was developed for application to general nucleosynthesis processes and is described in more detail in \cite{2016MNRAS.463.4153R}, where its first application was to the $\gamma$-process in massive stars. In the following we only provide an outline of the most important concepts and especially of details particular to $s$-process nucleosynthesis.

In this work we focus on reactions relevant to heavy element synthesis by the $s$-process. This involves nuclei with mass numbers $A>56$ and thus we do not vary reaction rates for lighter nuclei.

\subsubsection{Uncertainty of neutron capture rates}
\label{sec-ng}

Reaction rates in nucleosynthesis, even those experimentally determined under laboratory conditions, can bear significant theoretical uncertainty due to population of excited states at stellar temperatures. For the $s$-process, many neutron-capture rates based on experimental data are available, but \cite{2011ApJ...738..143R} and \cite{2012ApJS..201...26R} demonstrated that excited state contributions can be important even at $s$-process temperatures. Thus, we adopt temperature-dependent uncertainty factors, based on the contribution of reactions on the target ground state (as measured in the laboratory) to the reaction \citep{2012ApJS..201...26R}. Using the ground-state contribution $X_0$, we calculate the uncertainty of (n,\,$\gamma$)-reactions by
\begin{equation}
	\label{eq-unc_ng}
	u_{({\mbox n},\gamma)}(T) = u_{\rm exp} X_0(T) + u_{\rm th} \left( 1 - X_0(T) \right)  \ ,
\end{equation}
where $u_{\rm exp}$ and  $u_{\rm th}$ are the uncertainty factors for experiment and theory, respectively.

The value of $X_{0}$ behaves monotonically with temperature, approaching zero with increasing temperature. Thus, we obtain $u_{({\mbox n},\gamma)} \simeq u_{\rm exp}$ at low temperatures and $u_{({\mbox n},\gamma)} \simeq u_{\rm th}$  at high temperatures, respectively. Experimental uncertainties (2$\sigma$) are used for the measured ground state rates, whereas $u=2$ is adopted for unmeasured rates and for reactions on thermally exited states. We apply $u_{({\mbox n}, \gamma)}$ for the upper limit and ${u_{({\mbox n}, \gamma)}}^{-1}$ for the lower limit in the uniform MC variation. Here, in the context of the $s$-process, the majority of reactions are based on experimental data. For more details and the derivation of Equation~\ref{eq-unc_ng}, see \cite{2016MNRAS.463.4153R} and references therein.

\subsubsection{Uncertainty of $\beta$-decay rates}
\label{sec-beta}

Although most $\beta$-decay half-lives for nuclei relevant to the $s$-process are based on experimental data, the temperature dependence for these half-lives is not well known. We therefore apply an approach similar to that described above for $\beta$-decay rates, but based on partition functions to determine the importance of excited states. The uncertainty at low temperature ($T < 10^7$~K) corresponds to the one of measured decays. A uniform random distribution between the upper and lower limit of the reaction rate at a given temperature is used for the MC variation factors.

The temperature-dependent uncertainty for $\beta$-decay rates is given by
\begin{equation}
  u_{\beta}(T) = \frac{2J_0 + 1}{G(T)}u_{\rm g.s.} + \left( 1 - \frac{2J_0 + 1}{{G(T)}} \right) u_{\rm e.s.} 
\end{equation}
where $G(T)$ is the temperature dependent partition function \citep[see, {\it e.g.},][]{2000ADNDT..75....1R}. The value of $G$ generally reaches $2J_0 + 1$ at low temperature ($T_9 < 0.1$), {\it i.e.}, $u(T) \simeq u_{\rm g.s.}$, while $G$ becomes larger as the temperature increases. Thus, $\beta$-decay rates become more dependent on theory uncertainties with increasing temperature, which is due to the increasing contribution from excited state decays. In this study, we adopt $u_{\rm g.s.} = 1.3$ and $u_{\rm e.s.} = 10$ unless experimentally known. Nevertheless, the total uncertainty remains within a factor of few in the $s$-process temperature range.

In addition to $\beta$-decay, electron captures (${\rm e}^{-}$-captures) on nuclei are taken into account as provided by \cite{freiburghaus1999}. While ${\rm e}^{-}$-capture has less impact on the $s$-process as compared to $\beta$-decay, the uncertainty in its decay rates is large. In this study, we simply adopt a constant factor $2$ for variation of all ${\rm e}^{-}$-capture rates and do not study this feature in more detail.

\subsection{MC simulations}


We determine the upper and lower limits of each reaction rate as described above, and vary the reaction rate in each MC calculation step. We adopt a uniform distribution of values between the limits for this random variation. Since the uncertainty factors are not evaluated analytically, we fit the upper and lower rate limits for computational efficiency, using Equation~\ref{eq-reaclib}. We find that $10,000$ MC iterations gives well converged results \citep[see][for the $\gamma$-process that requires a much larger reaction network]{2016MNRAS.463.4153R}. 

In the simulations, all relevant rates are varied simultaneously within the assigned uncertainties. As we focus on $s$-process nucleosynthesis, we included all neutron captures and weak rates (mostly $\beta$-decays) for heavier nuclei beyond iron ($Z>26$) in the MC variation. This amounts to $900$ reactions being varied in total. For comparison, we also calculated cases with variation of only (n,$\gamma$) or only weak reactions. This included variations of $510$ and $390$ reactions, respectively. For every case, we performed $10,000$ MC iterations, required because convergence depends on the total number of rates in the reaction network and does not depend on the number of reactions varied \citep[see also][]{2016MNRAS.463.4153R}.

In each MC iteration, the rate $r_i$ of each reaction $i$ received its specific random variation factor $f_i$. The same factor was applied to the respective reverse rate. Although each initially assigned $f_i$ is a single, randomly determined value between $0$ and $1$, the actual rate variation factor is temperature dependent because of the temperature-dependent upper and lower limits of the uncertainty ranges. The initial factor $f_i$ is mapped consistently to an actual rate variation factor at each temperature \citep[see][for more details]{2016MNRAS.463.4153R}. It should be noted that although the relation between the variation factor and the rate is linear and monotonic, variation factors and final abundances have a strongly non-linear, and sometimes non-monotonic relation.

\begin{figure}
  \centering
  \includegraphics[width=\columnwidth]{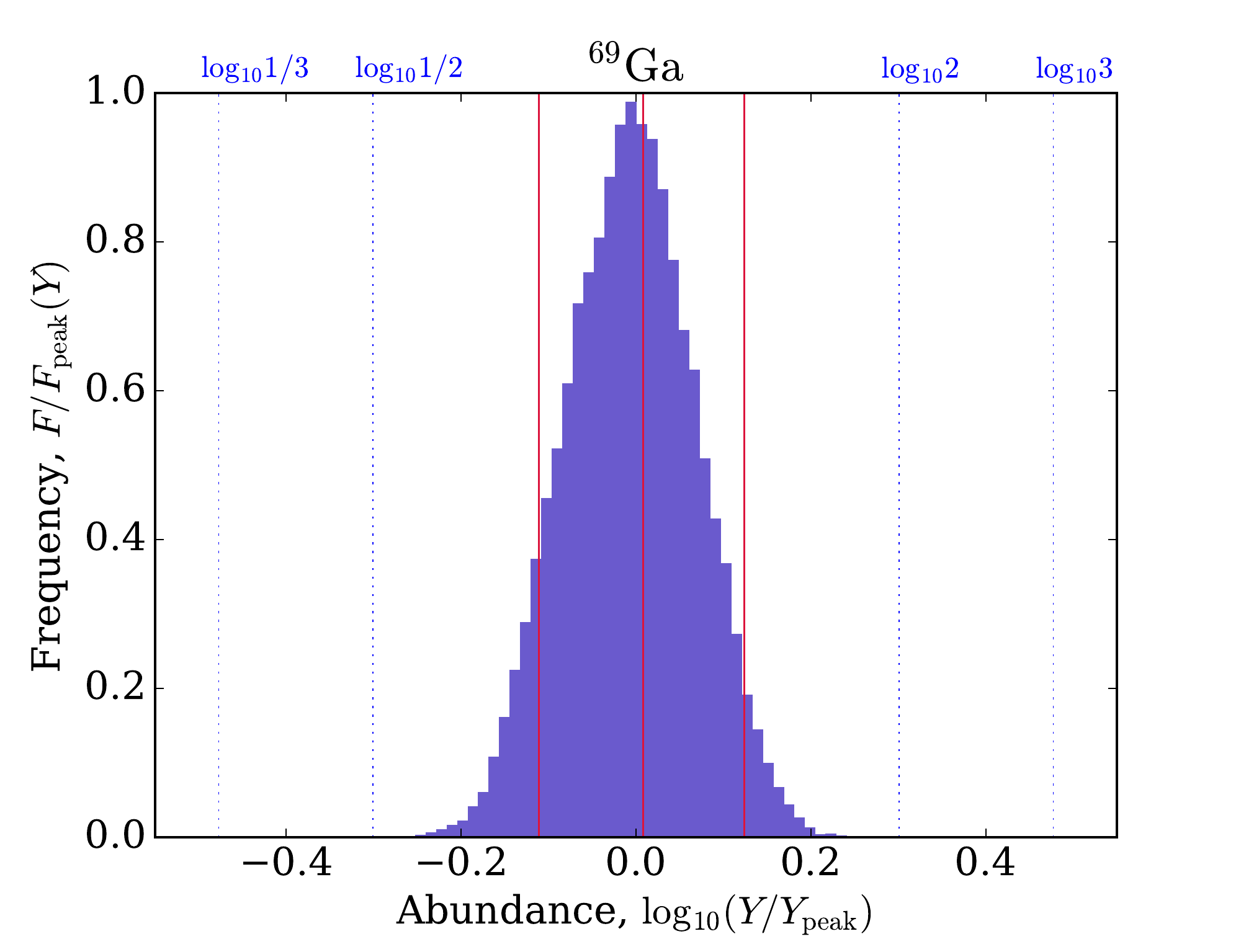}
  \includegraphics[width=\columnwidth]{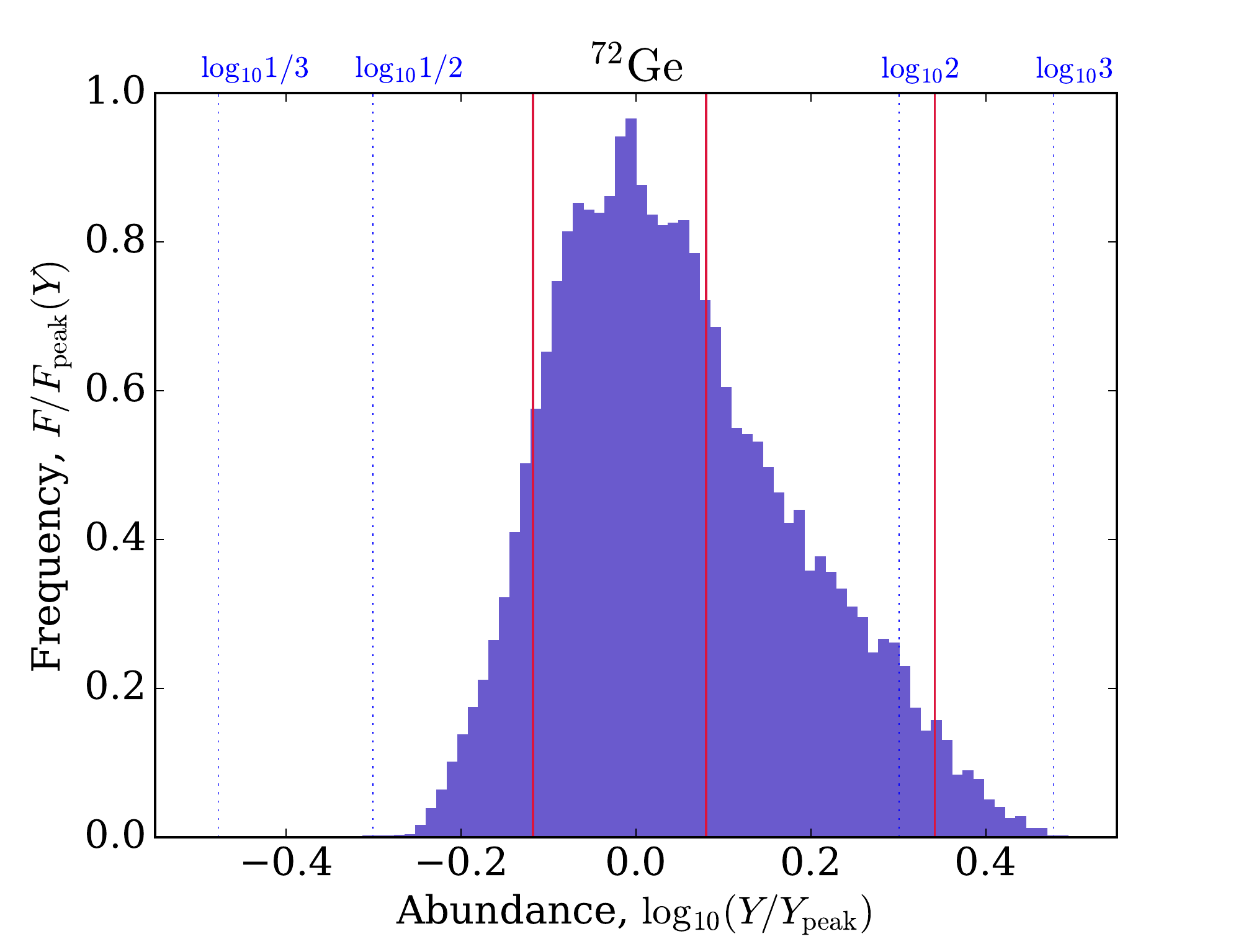}
  \caption{The final abundance distributions of $^{69}{\rm Ga}$ and $^{72}{\rm Ge}$ in the {\it ws}-process, based on the result of $10,000$ MC iterations. The plot shows the histogram of the frequency $F$ for the final abundance $Y$ normalized to the peak value, $F_{\rm peak}$. Red lines correspond to the values of $5\%$, $50\%$ and $95\%$ in the cumulative frequency. Note that the histogram is plotted for logarithmic value of abundances, {\it i.e.,} $\log_{10} Y/Y_{\rm peak}$, not for $Y/Y_{\rm peak}$.}
   \label{fig-mc-hist}
\end{figure}

\begin{figure}
  \centering
  \includegraphics[width=\columnwidth]{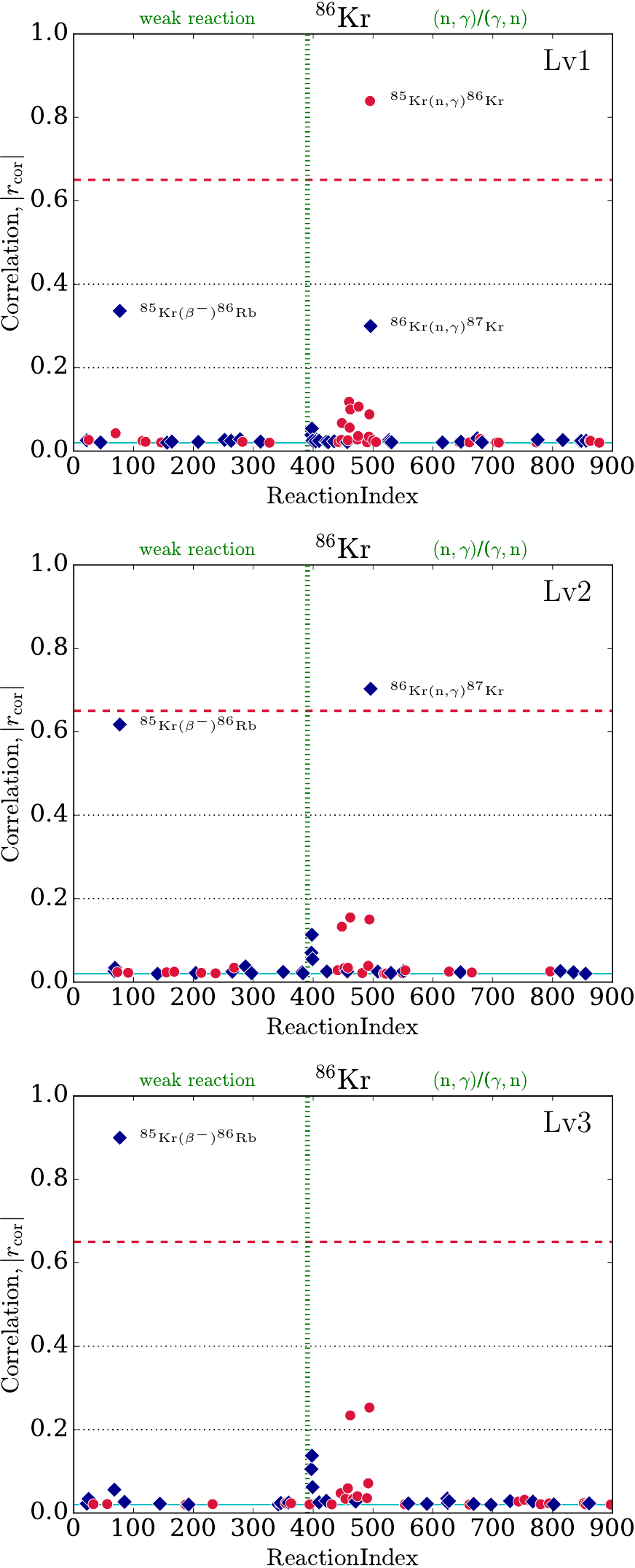}
  \caption{The correlation coefficients of reactions with respect to an abundance change of $\iso{Kr}{86}$ in the {\it ws}-process, obtained in MC calculations with reaction rate variations concerning \textit{Level 1} (Lv1, top), Lv2 (middle), and Lv3 (bottom) key rates. The absolute values of the coefficients are plotted against a reaction index number. Red circles stand for positive correlation and blue squares for negative correlation, respectively. Reaction indices in the range of $1$--$390$ denote weak reactions and those in the range $391$--$900$ identify neutron captures. Note that, for better readability, reactions with correlation factors $|r_{\rm cor}| < 0.02$ are omitted from this plot.}
   \label{fig-mc-cor}
\end{figure}

The result of each MC run consists of a set of final abundances for each isotope. Thus $10,000$ different sets of isotopic abundances were obtained for each variation case. As an example, Figure~\ref{fig-mc-hist} shows the frequency ($F$) of final abundance ($Y$) distributions of $\iso{Ga}{69}$ and $\iso{Ge}{72}$ in the {\it ws}-process (see, Section~\ref{sec-wsproc} for details). The peak value of the final abundance $Y$ is denoted by $Y_{\rm peak}$ and the frequency of abundance values found in the iterations is normalized to this value. Red lines indicate $5\%$ and $95\%$ of the cumulative frequency and thus the interval between the lines contains $90\%$ of the results (see {\it e.g.}, Figure~\ref{fig-mc-ws-all} and following). We adopt this interval as uncertainty in the final abundance. Note that each distribution is not exactly a Gaussian or lognormal distribution, although the histogram has a continuous shape.

The uncertainty of $\iso{Ga}{69}$ appears to be symmetrically distributed, {\it i.e.}, it is centered around $Y_{\rm peak} \simeq Y(50\%)$, and the uncertainty is significantly below a factor of two. On the other hand, $\iso{Ge}{72}$ has an asymmetric distribution with a longer tail at larger values. Resulting from this asymmetry, the peak of the distribution does not correspond to the average value of $Y$, {\it i.e.}, $Y_{\rm peak} \neq Y(50\%)$. The uncertainty range, determined by $Y(5\%)$ and $Y(95\%)$, exceeds by a factor of two, while the minimum limit is closer to one than to a factor of $1/2$.

\subsection{Key reaction rates based on MC calculations}
\label{sec-cor-coff}

As we obtain sets of rate variation factors and corresponding distributions of final abundances, the statistical correlation between them can be investigated. Key reactions are then identified by a strong correlation, as introduced in \cite{2016MNRAS.463.4153R}. In the current study, we calculate $900\times N_{\rm nuc}$ correlation factors (number of varied reactions $\times$ number of nuclei of interest: $N_{\rm nuc}$).

We adopt the Pearson product-moment correlation coefficient \cite{Pearson01011895} to quantify the correlation between rate variation and the final abundances \citep[also used in][]{2016MNRAS.463.4153R}, defined by
\begin{equation}
	r_{\rm cor} = \frac{\displaystyle \sum_{i}^{n}(x_i - \bar{x})(y_i - \bar{y})}
	{\displaystyle \sqrt{\sum_{i}^{n}(x_i - \bar{x})^2}\sqrt{\sum_{i}^{n}(y_i - \bar{y})^2}}
	\label{eq-cor}
\end{equation}
where $x_i$ and $y_i$ are variables with $\bar{x}$ and $\bar{y}$ being their arithmetic mean value, respectively. The summation is applied to all data for the MC runs $i=1, 2, 3, \cdots, n$. Here, $x$ and $y$ in Equation~\ref{eq-cor} correspond to variation factors $f$ and final abundances $Y$. 

The value $r_{\rm cor}$ ranges between $-1$ to $+1$ and the absolute value ($0 \leq |r_{\rm cor}| \leq 1$) indicates the correlation strength. Following our previous MC analysis for the $\gamma$-process \citep{2016MNRAS.463.4153R}, we assume $|r_{\rm cor}| > 0.7$ to be a strong correlation, whereas a value below $0.2$ indicates no correlation. As the correlation strength changes gradually and for numerical stability, we adopt $0.65$ as the threshold for a significant correlation value.

In Figure~\ref{fig-mc-cor}, we plot $|r_{\rm cor}|$ for $\iso{Kr}{86}$ in the {\it ws}-process (see Section~\ref{sec-wsproc} for details).
The top panel labeled ``Lv1'' corresponds to the results of a MC run varying all weak and (n,$\gamma$) reaction rates (\textit{Level 1}). The name of a reaction is indicated for a few cases with higher $|r_{\rm cor}|$, which are, for example, $\iso{Kr}{85}({\mbox n}, \gamma)\iso{Kr}{86}$ producing $\iso{Kr}{86}$ and $\iso{Kr}{86}({\mbox n}, \gamma)\iso{Kr}{87}$ destroying $\iso{Kr}{86}$.

For the production of $\iso{Kr}{86}$, we find two neutron capture reactions with elevated correlation values. Only one of them, $\iso{Kr}{85}({\mbox n}, \gamma)\iso{Kr}{86}$, has a significantly high correlation of $0.8$. The correlation factor of the other one, $\iso{Kr}{86}({\mbox n}, \gamma)\iso{Kr}{87}$, is around $0.3$. Among the weak rates, the largest correlation factor is found for $\iso{Kr}{85}(\beta^{-})\iso{Rb}{85}$, with $|r_{\rm cor}| = 0.2$. Since only values $|r_{\rm cor}| \ge 0.65$ can be considered to be a strong correlation, only $\iso{Kr}{85}({\mbox n}, \gamma)\iso{Kr}{86}$ is chosen as a key reaction rate regarding abundance changes of $\iso{Kr}{86}$. In such a manner key rates were identified for all investigated nuclides.

Key rates found in the first MC run varying all rates simultaneously are labeled \textit{Level 1} key rates and the corresponding MC run is the Lv1 MC run. Following \cite{2016MNRAS.463.4153R}, we also investigated further (lower) levels of key reactions. To see how the final uncertainties are reduced when Lv1 key reactions are determined (by future measurements or theoretical predictions), further MC runs were performed excluding these from the MC rate variation. This defines a Lv2 MC calculation. As shown in Figure~\ref{fig-mc-cor}~(middle), the correlation values of the remaining varied reactions is expected to increase compared to the Lv1 run, because the most dominant reactions are no longer varied, and thus are not considered, in the Lv2 calculation. On the other hand, it is obvious that the resulting uncertainties in the final abundances obtained in the Lv2 MC run are decreased with respect to those from the previous calculation. Based on the results of Lv2 MC run, we also identified additional key reaction rates, the Lv2 key rates, using the same criterion for the correlation value as before.

Another iteration of the same screening method was used to find Lv3 key reaction rates after having performed the Lv3 MC calculation, which implies that Lv1 and Lv2 key reactions are set to the standard reaction rate and not varied. As shown in Figure~\ref{fig-mc-cor}~(lower), finally the correlation for $\iso{Kr}{85}(\beta^{-})\iso{Rb}{85}$ exceeds $r_{\rm cor} = 0.65$ and thus it becomes a Lv3 key rate. In the following, we show key reactions at various levels for the {\it ws}-process and the {\it es}-process, described in Sections~\ref{sec-wsproc} and \ref{sec-esproc}, respectively.

An important point becomes obvious from the above: Lv2 and Lv3 key rates become important only \textit{after the uncertainties for all key reactions in the higher levels have been reduced}. An improved constraint of a Lv2 or Lv3 rate will have no significant impact if key rates of higher levels are still only weakly constrained. Nevertheless, providing also Lv2 and Lv3 key rates may be useful to determine long-term research strategies.

The Lv2 (and Lv3) key rates identified here incur an additional uncertainty. The methodology used identifies these rates under the assumption that Lv1 (Lv2 rates) are constrained at their standard values with their existing uncertainty. If new measurements were to constrain a Lv1 (Lv2) key rate at a new, different value, with different uncertainties, then the redetermination of Lv2 and Lv3 key reactions would be in order.

\begin{figure}
  \centering
  \includegraphics[width=\hsize]{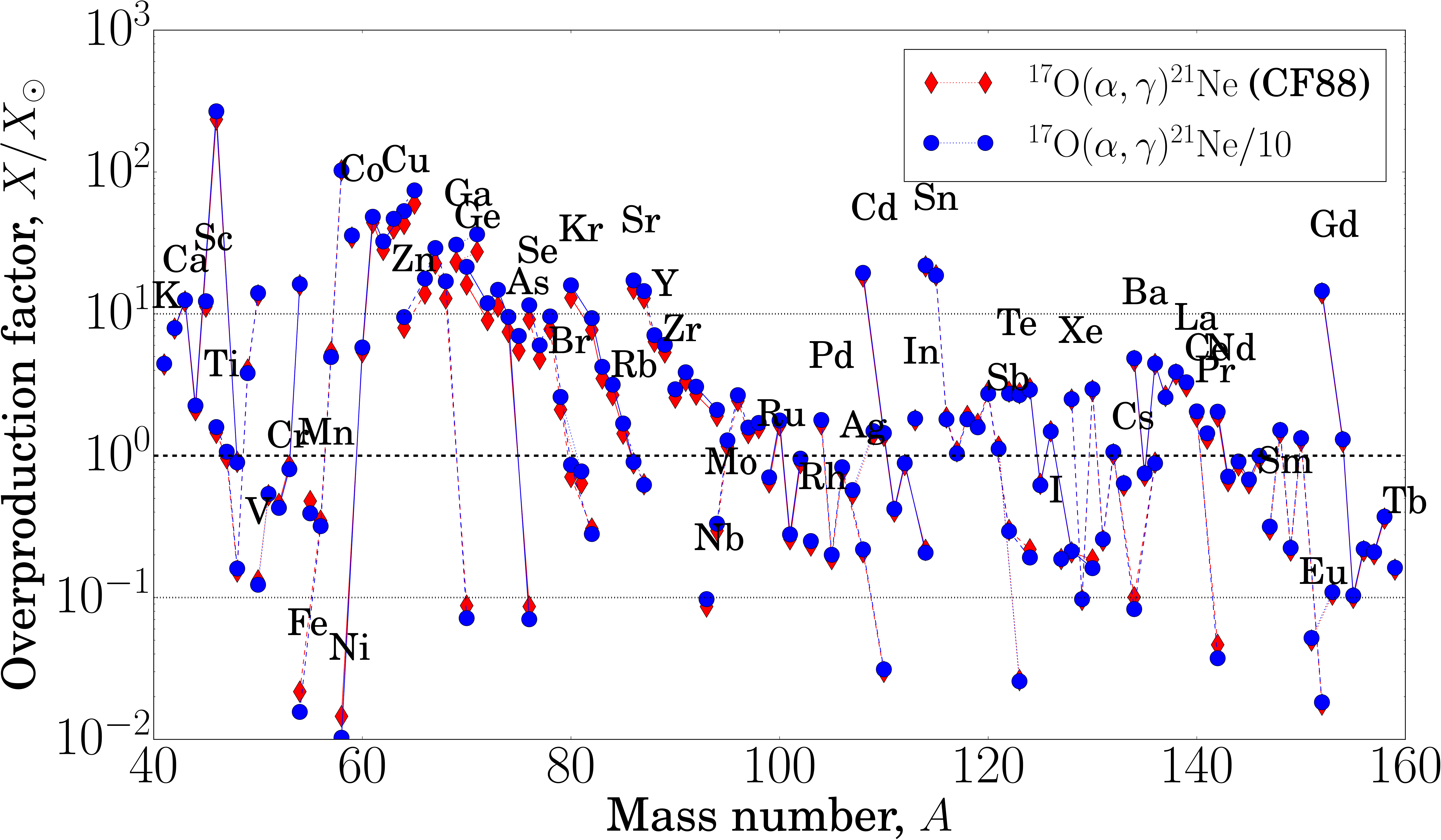}
  \caption{Final overproduction factors of the {\it ws}-process, based on the {\tt z0r0} model. Nucleosynthesis results using two rates for  $^{17}{\rm O} (\alpha, \gamma)^{21}{\rm Ne}$ are shown: for the standard rate by CF88 (red diamonds) and the CF88 rate divided by 10 (blue circles).}
  \label{fig-fabund_ws}
\end{figure}

\begin{figure}
  \centering
    \includegraphics[width=\hsize]{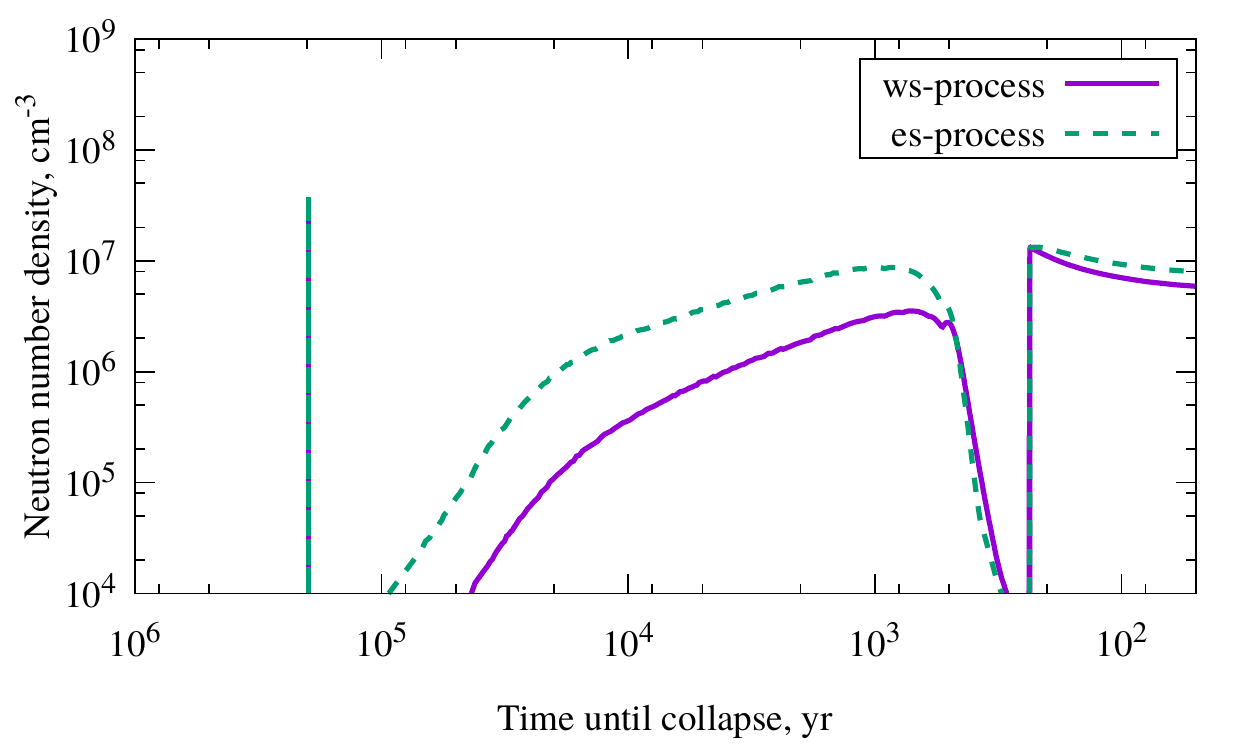}
  \caption{The evolution of the neutron number density for the {\it ws}-process (solid line) and {\it es}-process (dashed line). See Figure~\ref{fig-star_hydro} for the corresponding density and temperature evolution.}
     \label{fig-ndens}
\end{figure}

\begin{figure}
  \centering
    \includegraphics[width=\hsize]{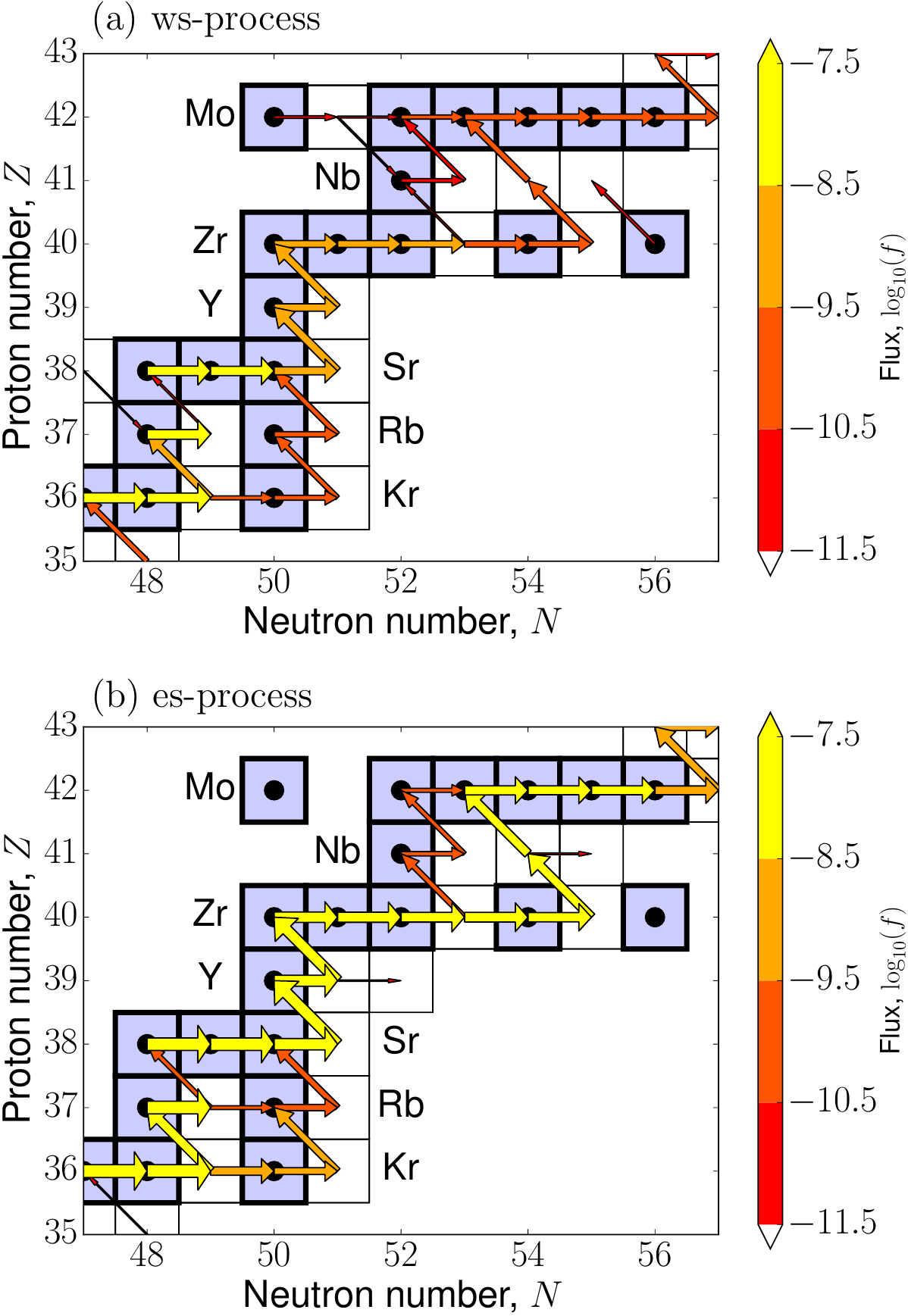}
  \caption{Nucleosynthesis flows in the (a) {\it ws}-process and (b) {\it es}-process. The time-integrated abundance change (flux) of each reaction is shown by an arrow. The flux value is indicated by the colour and width of the arrow.}
     \label{fig-flow_ws}
\end{figure}

\section{The weak \lowercase{$s$}-process}
\label{sec-wsproc}

\subsection{Nucleosynthetic features of the {\it ws}-process}
\label{sec-ws-basis}

The final abundance distribution of the {\it ws}-process, {\it i.e.} just before the onset of the core collapse, is shown in Figure~\ref{fig-fabund_ws}. We chose a solar-metallicity star without rotation, the {\tt z0r0} model, as the standard case for the {\it ws}-process. Two different abundance distributions are plotted, based on different reaction rates for the neutron-poison relevant reaction $\iso{O}{17}(\alpha,\gamma)\iso{Ne}{21}$, which has a significant physical uncertainty. We used the rate adopted by \cite{1988ADNDT..40..283C}, denoted CF88, and a rate value divided by a factor of $10$ to consider a recent experimental suggestion for the associated uncertainty (as discussed in Section~\ref{sec-network}). Despite the difference in these rate values, the abundance patterns are almost identical at $Z_{\rm m} = Z_\odot$ because the impact of the neutron poison $\iso{O}{16}$ is significant only for metal-poor stars \citep{2016MNRAS.456.1803F}. The abundance distribution agrees with a typical {\it ws}-process pattern \citep[see, {\it e.g.}][]{1990A&A...234..211P, 2016MNRAS.456.1803F}, of which the overproduction peak is at $A \simeq 60$ and the production steeply declines for nuclei with mass numbers $A \geq 90$.

The time evolution of the neutron density is shown in Figure~\ref{fig-ndens}, where the solid line corresponds to the {\it ws}-process result. As expected from the temperature and density evolution (shown in Figure~\ref{fig-star_hydro}), the neutron density has a peak at the beginning of the core-helium burning phase ($2 \times 10^5$ yr before collapse), exceeding $10^7 {\rm cm}^{-3}$ for a very short period. This increase is due to the $\iso{C}{13}(\alpha, {\mbox n})\iso{O}{16}$ reaction, which also is a dominant neutron source reaction for the main $s$-process in low mass asymptotic-giant-branch stars. However, in the evolution of a massive star, the duration of this peak is so short ($\sim 10$ yr) that this increase of the neutron density has no significant impact on the total neutron exposure. During the core-helium burning phase ($\sim 10^5$--$10^{3}$ yr before collapse), the neutron density assumes values $> 10^{5} {\rm cm}^{-3}$. After the ignition of carbon-shell burning at $\sim 3 \times 10^{2}$ yr before collapse, the neutron density increases further, although the duration of this phase is shorter than the core-He burning phase. The $\iso{O}{17}(\alpha,\gamma)\iso{Ne}{21}$ rate does not change the results significantly, so the reduced rate (CF88 divided by a factor of $10$) has been adopted for consistency with the {\it es}-process calculations (see later).

The nucleosynthesis flux of each reaction, {\it i.e.}, (n,$\gamma$) reactions and $\beta$-decays, over the nucleosynthesis time has been calculated. This 
equates to the time-integrated abundance change of each reaction from its initial abundance to its final abundance. Nucleosynthesis fluxes in the {\it ws}- and {\it es}-process obtained in this manner are shown in Figure~\ref{fig-flow_ws}. The colour and width of an arrow indicate the value of bulk flow (abundance change) for individual reactions. Note that the value of the nuclear flow is integrated over the entire nucleosynthesis calculation, which is different from the reaction rate at a given time step.

As expected for the $s$-process, the predominant reactions in nucleosynthesis are (n,$\gamma$) reactions ($\rightarrow$) and $\beta^{-}$ decays ($\nwarrow$) along the line of stability. Although the reaction flow is basically a single path, several branches are evident where a decay rate is comparable to an (n,$\gamma$) rate. This is the case for example at neutron numbers $N = 49$ and $53$ in the region plotted. For these branching points, we expect that the effect of nuclear physics uncertainty on the final abundances is more complicated due to the competition between neutron capture and $\beta$-decay. Note that for weak reactions in the $s$-process, $e^{-}$-capture also contributes to the reaction flow, {\it i.e.}, diagonal arrows in the plot. However, $\beta^-$-decay has a more significant impact on nucleosynthesis compared to the corresponding $e^{-}$-capture.

\begin{figure}
  \centering
  \includegraphics[width=\columnwidth]{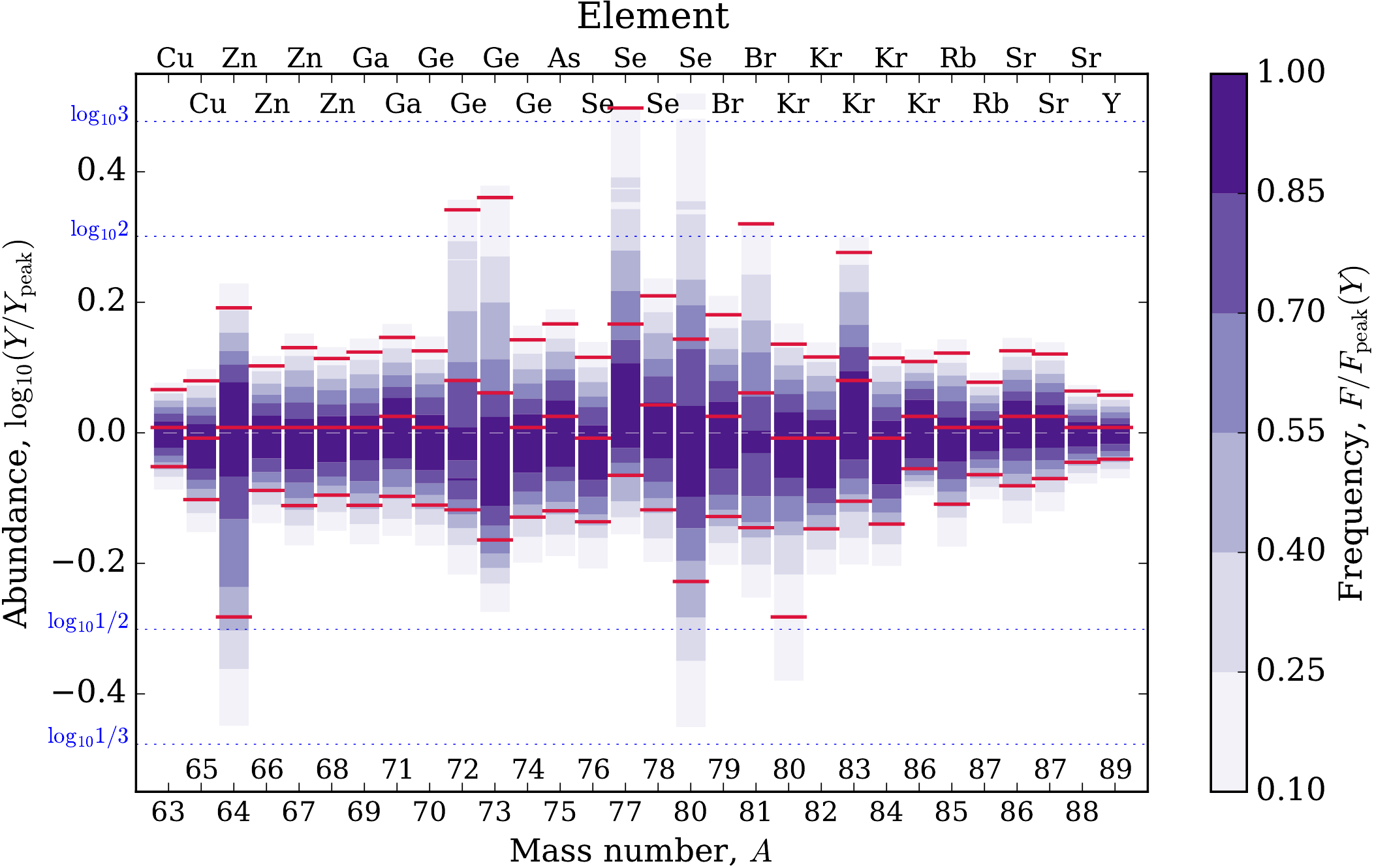}
  \caption{Uncertainty distributions for the {\it ws}-process when varying all neutron captures and weak rates. The colour shade is the probabilistic frequency and the 90\% probability intervals up and down marked for each nuclide (see, Figure~\ref{fig-mc-hist} for examples of the distribution). Horizontal dashed lines indicate uncertainty factors of $2$, $3$, $1/2$, and $1/3$, respectively.}
   \label{fig-mc-ws-all}
\end{figure}

\begin{figure}
  \centering
  \includegraphics[width=\columnwidth]{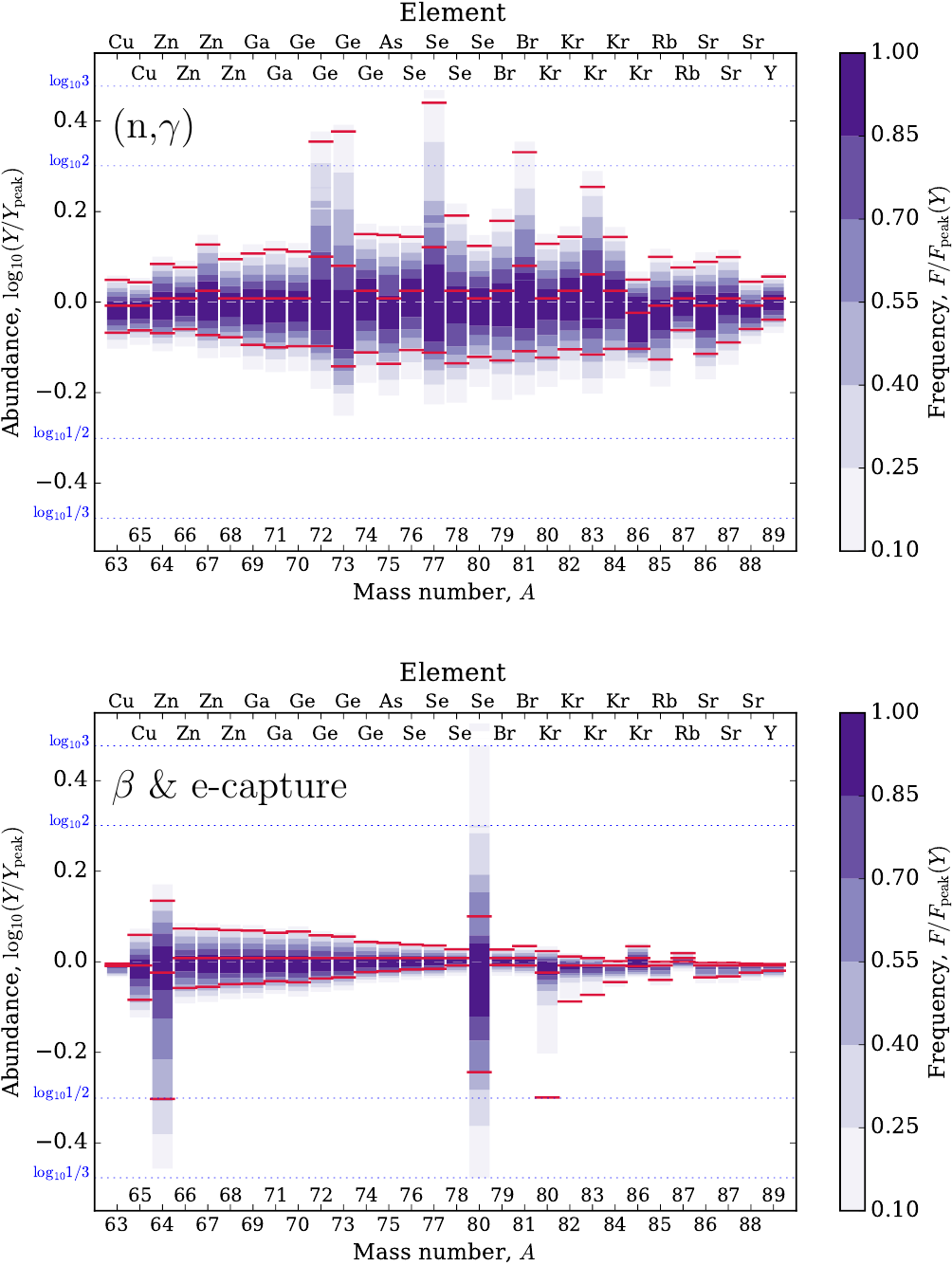}
  \caption{Same as Figure~\ref{fig-mc-ws-all},  varying only (n,$\gamma$) reactions (top) or weak rates (bottom).}
   \label{fig-mc-ws-only}
\end{figure}

\begin{table}
\centering
\caption{ \label{tab-ws-unc} Uncertainties in the final abundance of {\it ws}-process nuclei from the MC calculation. The column labeled ``Level'' indicates the level of the first key reaction found, as described in Section~\ref{sec-ws-key}. The remaining columns show uncertainty factors for variations Up and Down, of which values are $Y(95\%)/Y_{\rm peak}$ and $Y(5\%)/Y_{\rm peak}$, respectively. They enclose a 90\% probability interval, as shown in Figure~\ref{fig-mc-ws-all}. }
\begin{tabular}{@{}lcrr}
\hline
 & Level & Up & Down \\
\hline
${}^{63}{\rm Cu}$ & -- &  1.16 & 0.888 \\
${}^{65}{\rm Cu}$ & -- &  1.20 & 0.790 \\
${}^{64}{\rm Zn}$ &  1 &  1.55 & 0.522 \\
${}^{66}{\rm Zn}$ & -- &  1.27 & 0.816 \\
${}^{67}{\rm Zn}$ &  1 &  1.35 & 0.773 \\
${}^{68}{\rm Zn}$ & -- &  1.30 & 0.802 \\
${}^{69}{\rm Ga}$ & -- &  1.33 & 0.774 \\
${}^{71}{\rm Ga}$ & -- &  1.40 & 0.799 \\
${}^{70}{\rm Ge}$ & -- &  1.33 & 0.775 \\
${}^{72}{\rm Ge}$ &  1 &  2.20 & 0.762 \\
${}^{73}{\rm Ge}$ &  1 &  2.29 & 0.685 \\
${}^{74}{\rm Ge}$ &  3 &  1.39 & 0.743 \\
${}^{75}{\rm As}$ &  3 &  1.47 & 0.759 \\
${}^{76}{\rm Se}$ & -- &  1.31 & 0.731 \\
${}^{77}{\rm Se}$ &  1 &  3.15 & 0.861 \\
${}^{78}{\rm Se}$ &  1 &  1.62 & 0.762 \\
${}^{80}{\rm Se}$ &  1 &  4.61 & 0.592 \\
${}^{79}{\rm Br}$ &  2 &  1.52 & 0.744 \\
${}^{81}{\rm Br}$ &  1 &  2.09 & 0.715 \\
${}^{80}{\rm Kr}$ & -- &  1.37 & 0.522 \\
${}^{82}{\rm Kr}$ & -- &  1.31 & 0.713 \\
${}^{83}{\rm Kr}$ &  1 &  1.89 & 0.785 \\
${}^{84}{\rm Kr}$ &  3 &  1.30 & 0.725 \\
${}^{86}{\rm Kr}$ &  1 &  1.29 & 0.881 \\
${}^{85}{\rm Rb}$ & -- &  1.33 & 0.778 \\
${}^{87}{\rm Rb}$ &  3 &  1.20 & 0.863 \\
${}^{86}{\rm Sr}$ & -- &  1.34 & 0.830 \\
${}^{87}{\rm Sr}$ & -- &  1.32 & 0.851 \\
${}^{88}{\rm Sr}$ & -- &  1.16 & 0.901 \\
${}^{89}{\rm Y}$ & -- &  1.14 & 0.911 \\
\hline
\end{tabular}
\medskip
\end{table}


\begin{figure}
  \centering
  \includegraphics[width=\columnwidth]{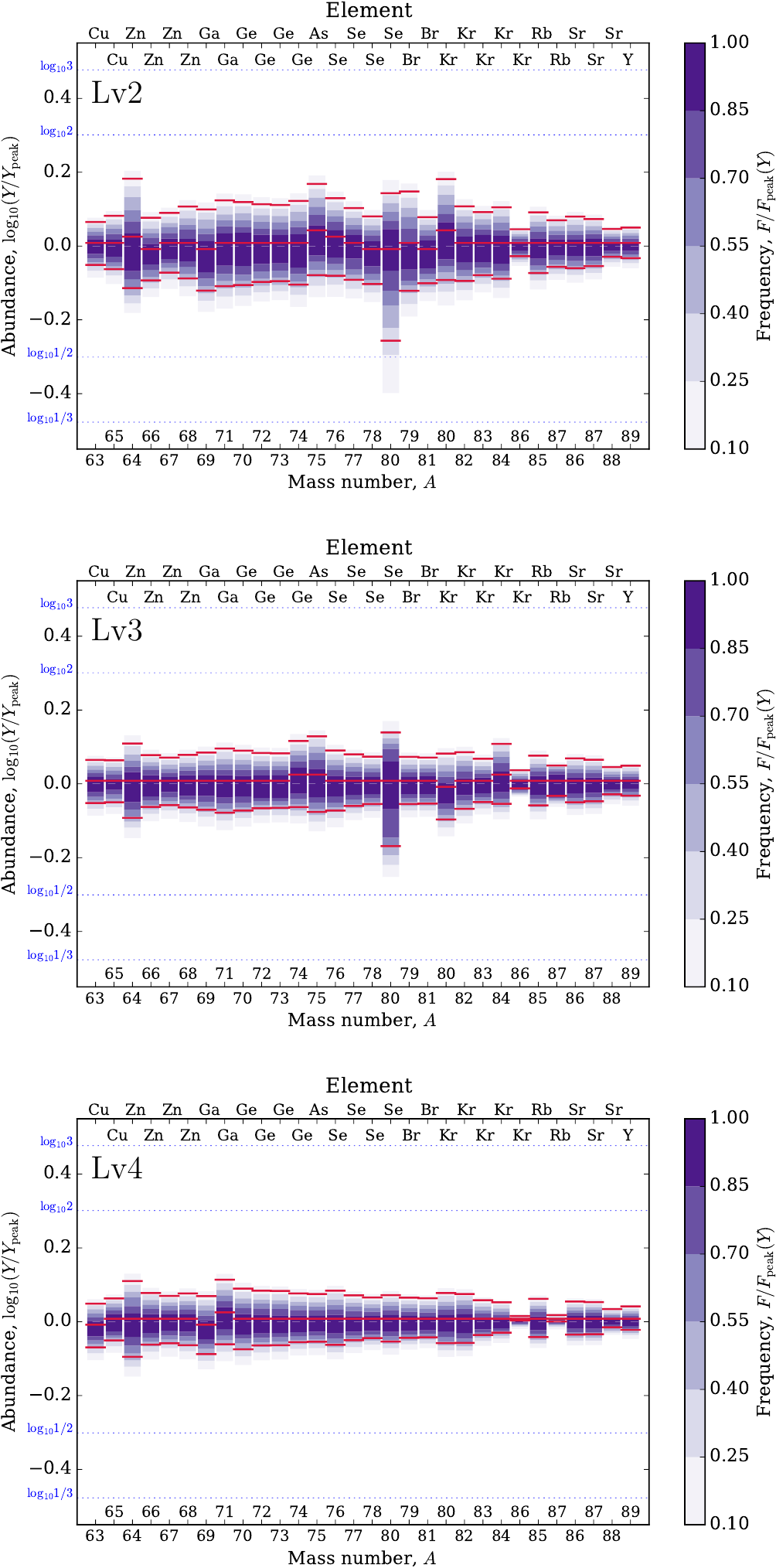}
  \caption{Results of MC calculations of the {\it ws}-process at different levels, Lv2 (upper panel), Lv3 (middle panel) and Lv4 (lower panel). Uncertainty distributions are plotted in the same manner as in Figure~\ref{fig-mc-ws-all}.}
   \label{fig-mc-ws-lv}
\end{figure}

\begin{table*}
\centering
\caption{ \label{tab-ws-key}The key reaction rates for the {\it ws}-process. Key rates in levels 1--3 are shown, along with their correlation factors $r_{\rm cor,0}$, $r_{\rm cor,1}$ and $r_{\rm cor,2}$, respectively. Significant correlation values are underlined. Not all $s$-process nuclei are listed but only those for which key rates were found. Also shown for each rate are the g.s. contributions of the (n,$\gamma$) reaction to the stellar rate and uncertainty factors of the $\beta$-decay rate at two plasma temperatures, respectively.}
\begin{tabular}{@{}l|c|ccccccccc}
\hline
Nuclide & $r_{{\rm cor},0}$ & $r_{{\rm cor},1}$ & $r_{{\rm cor},2}$& Key Rate & Key Rate & Key Rate& $X_0$ & Weak Rate \\
&&&& Level 1 & Level 2 & Level 3 & $(8$, $30~{\rm keV})$ & $(8$, $30~{\rm keV})$\\
\hline
$\iso{Zn}{64}$ & \underline{ 0.76} &       &       & $\iso{Cu}{64}(\beta^-)\iso{Zn}{64}$& $ $& $ $& & 1.30, 1.36\\
     & -0.46 & \underline{-0.73} &       & $ $& $\iso{Cu}{64}({\rm e}^-,\nu_{\rm e})\iso{Ni}{64}$& $ $& & ${\rm e}^-$ capture\\
$\iso{Zn}{67}$ & \underline{-0.67} &       &       & $\iso{Zn}{67}(\mbox{n},\gamma)\iso{Zn}{68}$& $ $& $ $&1.00, 1.00&\\
$\iso{Ge}{72}$ & \underline{-0.85} &       &       & $\iso{Ge}{72}(\mbox{n},\gamma)\iso{Ge}{73}$& $ $& $ $&1.00, 1.00&\\
$\iso{Ge}{73}$ & \underline{-0.84} &       &       & $\iso{Ge}{73}(\mbox{n},\gamma)\iso{Ge}{74}$& $ $& $ $&0.88, 0.81&\\
$\iso{Ge}{74}$ & -0.44 & -0.54 & \underline{-0.67} & $ $& $ $& $\iso{Ge}{74}(\mbox{n},\gamma)\iso{Ge}{75}$&1.00, 1.00&\\
$\iso{As}{75}$ & -0.50 & -0.59 & \underline{-0.70} & $ $& $ $& $\iso{As}{75}(\mbox{n},\gamma)\iso{As}{76}$&1.00, 1.00&\\
$\iso{Se}{77}$ & \underline{-0.86} &       &       & $\iso{Se}{77}(\mbox{n},\gamma)\iso{Se}{78}$& $ $& $ $&1.00, 1.00&\\
$\iso{Se}{78}$ & \underline{-0.71} &       &       & $\iso{Se}{78}(\mbox{n},\gamma)\iso{Se}{79}$& $ $& $ $&1.00, 1.00&\\
     &  0.38 & \underline{ 0.68} &       & $ $& $\iso{Zn}{68}(\mbox{n},\gamma)\iso{Zn}{69}$& $ $&1.00, 1.00&\\
$\iso{Se}{80}$ & \underline{-0.76} &       &       & $\iso{Br}{80}(\beta^-)\iso{Kr}{80}$& $ $& $ $& & 1.31, 4.70\\
     &  0.27 & \underline{ 0.73} &       & $ $& $\iso{Br}{80}(\beta^+)\iso{Se}{80}$& $ $& & 1.31, 4.70\\
     &  0.16 &  0.44 & \underline{ 0.88} & $ $& $ $& $\iso{Br}{80}({\rm e}^-,\nu_{\rm e})\iso{Se}{80}$& & ${\rm e}^-$ capture\\
$\iso{Br}{79}$ & -0.64 & \underline{-0.73} &       & $ $& $\iso{Br}{79}(\mbox{n},\gamma)\iso{Br}{80}$& $ $&1.00, 1.00&\\
$\iso{Br}{81}$ & \underline{-0.80} &       &       & $\iso{Kr}{81}(\mbox{n},\gamma)\iso{Kr}{82}$& $ $& $ $&1.00, 0.98&\\
$\iso{Kr}{83}$ & \underline{-0.76} &       &       & $\iso{Kr}{83}(\mbox{n},\gamma)\iso{Kr}{84}$& $ $& $ $&0.81, 0.74&\\
$\iso{Kr}{84}$ & -0.49 & -0.65 & \underline{-0.76} & $ $& $ $& $\iso{Kr}{84}(\mbox{n},\gamma)\iso{Kr}{85}$&1.00, 1.00&\\
$\iso{Kr}{86}$ & \underline{ 0.84} &       &       & $\iso{Kr}{85}(\mbox{n},\gamma)\iso{Kr}{86}$& $ $& $ $&1.00, 1.00&\\
     & -0.30 & \underline{-0.70} &       & $ $& $\iso{Kr}{86}(\mbox{n},\gamma)\iso{Kr}{87}$& $ $&1.00, 1.00&\\
     & -0.34 & -0.62 & \underline{-0.90} & $ $& $ $& $\iso{Kr}{85}(\beta^-)\iso{Rb}{85}$& & 1.30, 1.30\\
$\iso{Rb}{87}$ & -0.56 & -0.65 & \underline{-0.95} & $ $& $ $& $\iso{Rb}{87}(\mbox{n},\gamma)\iso{Rb}{88}$&1.00, 1.00&\\
\hline
\end{tabular}
\end{table*}

\begin{table}
\centering
\caption{ \label{tab-ws-unc-z} Uncertainties in the {\it ws}-process for elemental abundances. The columns `Up' and `Down' correspond to the upper and lower boundary of the uncertainty range similar to Table~\ref{tab-ws-unc} but for each element. The correlation coefficient $r_{\rm cor}$ and the corresponding reaction is shown for Lv1 key reactions (See Section~\ref{sec-ws-key}).}
\begin{tabular}{@{}ccccc}
\hline
Element & Up & Down & $r_{\rm cor}$ & Lv1 Key Reaction \\
\hline
  Cu &  1.16 & 0.891 &        & $                              $ \\
  Zn &  1.27 & 0.720 &  0.68 & $\iso{Cu}{64}(\beta^{-})\iso{Zn}{64}$ \\
  Ga &  1.33 & 0.778 &        & $                              $ \\
  Ge &  1.27 & 0.754 &        & $                              $ \\
  As &  1.47 & 0.759 &        & $                              $ \\
  Se &  1.40 & 0.737 &        & $                              $ \\
  Br &  1.57 & 0.732 &        & $                              $ \\
  Kr &  1.27 & 0.733 &        & $                              $ \\
  Rb &  1.29 & 0.804 &        & $                              $ \\
  Sr &  1.19 & 0.876 &        & $                              $ \\
   Y &  1.14 & 0.911 &        & $                              $ \\
\hline
\end{tabular}
\medskip
\end{table}

\subsection{Nuclear uncertainties in the {\it ws}-process}
\label{sec-mc-ws}

MC calculations for the {\it ws}-process have been performed, based on the rate variation method for neutron captures and weak rates as described in Section~\ref{sec-pizbuin}. The abundance uncertainty distributions for all {\it ws}-process nuclei are shown in Figure~\ref{fig-mc-ws-all} using the standard {\it ws}-process model {\tt z0r0} (see Section~\ref{sec-ws-basis}). The colour shade in the plot shows the frequency $F$ of each abundance $Y$ normalized to $F(Y_{\rm peak})$ as explained for Figure~\ref{fig-mc-hist}. Again, the interval between the red lines corresponds to 90\% of all abundance values. The numerical uncertainty value for each investigated nucleus is given in Table~\ref{tab-ws-unc}, in which the columns `Up' and `Down' correspond to the $Y(95\%)/Y_{\rm peak}$ and $Y(5\%)/Y_{\rm peak}$ values, respectively. The column `Level' in the table indicates the level of a key reaction relevant to the production or destruction of the nucleus, defined in Section~\ref{sec-cor-coff} and discussed in more detail in the following section.

As can be seen in Figure~\ref{fig-mc-ws-all} and Table~\ref{tab-ws-unc}, for most nuclides the uncertainty distributes symmetrically and the boundaries of the uncertainty range (90\% of cumulative frequency around the $Y_{\rm peak}$) are located at $F/F_{\rm peak} > 0.1$. We find that the uncertainty of most isotopes is smaller than a factor of two. Only a few species, specifically $\iso{Zn}{64}$, $\iso{Ge}{72,73}$, $\iso{Se}{77, 80}$, $\iso{Br}{81}$ and $\iso{Kr}{83}$, show a larger uncertainty. As already seen in Figure~\ref{fig-mc-hist}, the distribution is not symmetric for such nuclei, having a very much larger upper value or a very much smaller lower value, compared to the other boundary. Excepting these specific isotopes, the general trend in final abundance uncertainty is to increase from about 10\% at $A \sim 63$ to about $50$\% at $A \sim 80$. This reflects the propagation of uncertainties as the nucleosynthesis flow builds heavier nuclei from lighter nuclei. Above the mass number $A = 80$, the absence any reaction rates with significant uncertainties results in overall abundance uncertainties that then reduce with increasing mass.

To investigate the impact of uncertainties in neutron captures and weak reactions separately, we also performed MC calculations varying those rates separately. The results are shown in Figure~\ref{fig-mc-ws-only}. As the global feature of the uncertainty distribution for the (n,$\gamma$) variation case is similar to the results of varying all (n,$\gamma$) and weak reactions, it is demonstrated that the total uncertainty is mostly caused by the neutron captures, while weak reactions only have a minor contribution. Only for $\iso{Zn}{64}$ and $\iso{Se}{80}$ are the uncertainties dominated by weak reactions rather than by (n,$\gamma$) reactions. These isotopes are at a known $s$-process branching point.

\subsection{Key reactions for the weak $s$-process}
\label{sec-ws-key}

The correlation coefficients $r_{\rm cor}$ for all $s$-process isotopes and all neutron capture and weak rates were computed according to Equation~\ref{eq-cor} and used to identify key reactions mainly contributing to the abundance uncertainty of each isotope, as explained in Section~\ref{sec-cor-coff}. Only ten reactions bear a strong correlation $|r_{\rm cor}| \geq 0.65$ with final abundances. These key reactions are listed in Table~\ref{tab-ws-key}. The table has additional columns for key reactions at lower levels but the primary key reactions are listed in in the column ``Key Rate Level 1'' and its corresponding correlation coefficient is given in the column ``$r_{{\rm cor},0}$''.

As expected, most of the key reactions are neutron captures in the $s$-process path. A few weak reactions have significant impact for nuclei around branching points. The results of the MC runs at different levels are shown in Figure~\ref{fig-mc-ws-lv}. When the number of the reactions varied in the MC runs is decreased, the final uncertainties become smaller. The result of the Lv4 MC run shows a tiny uncertainty for all {\it ws}-process nuclei.

For reference, in Table~\ref{tab-ws-unc-z} we also provide uncertainty ranges and key correlations for \textit{elemental} abundances. The uncertainty range of each element is the weighted average value of the ones of its isotopes. All elements show uncertainty factors less than $\sim 1.5$ with the exception of Br, the upper limit of which is $1.57$. This is the case although some of the contributing isotopes of Ge, Se, and Br (more specifically $\iso{Ge}{72, 73}$, $\iso{Se}{77, 80}$ and $\iso{Br}{81}$) have a larger uncertainty beyond a factor of $2$ (see Table~\ref{tab-ws-unc}). Regarding key reactions for elemental abundances, only one case was found, the $\beta^{-}$ decay of $\iso{Cu}{64}$ which affects the production of Zn. This behaviour reflects that the production of each element involves contributions from multiple individual isotopes, that do not act coherently.


\section{The enhanced \lowercase{s}-process}
\label{sec-esproc}

\begin{figure}
  \centering
  \includegraphics[width=\hsize]{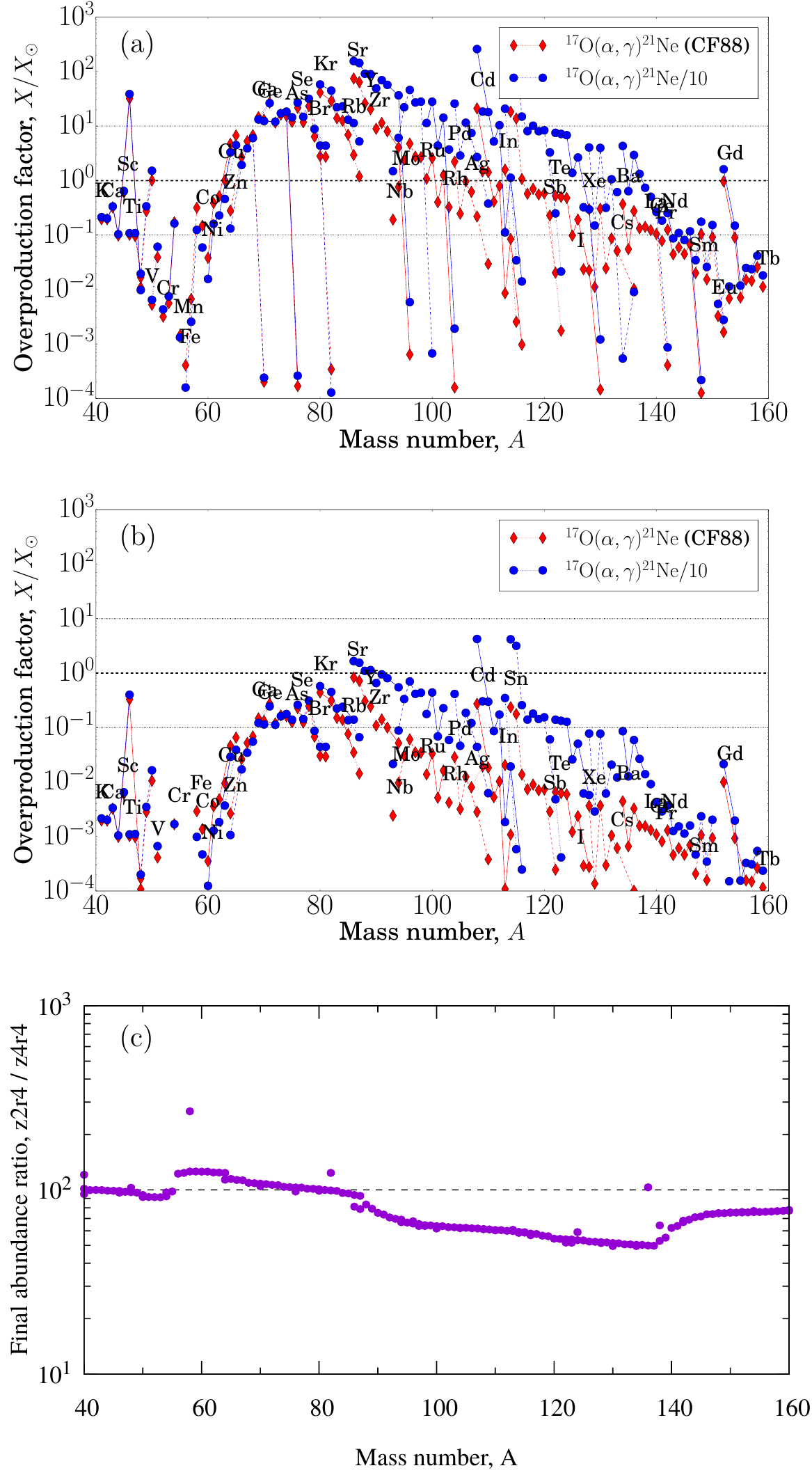}
  \caption{Overproduction factors in the {\it es}-process for (a) the {\tt z2r4} model and (b) the {\tt z4r4} model. Similarly to Figure~\ref{fig-fabund_ws}, results for different values of the $\iso{O}{17} (\alpha, \gamma)\iso{Ne}{21}$ rate are compared. (c) The final abundance ratio for the above models, based on a reduced $\iso{O}{17} (\alpha, \gamma)\iso{Ne}{21}$ reaction.}
  \label{fig-fabund_es}
\end{figure}

\subsection{Nucleosynthetic features of the {\it es}-process}
\label{sec-es-basis}

The results of nucleosynthesis calculations for rotating metal-poor stars are shown in Figure~\ref{fig-fabund_es}, for (a) the {\tt z2r4} model and (b) the {\tt z4r4} model. In both cases, the effect of rotation is included by adding 1\% by mass fraction of $^{14}$N to the initial composition. The {\tt z2r4} model is a low metallicity star with $Z_{\rm m} = 1 \times 10^{-3}$ and the {\tt z4r4} model is an even lower metallicity star with $Z_{\rm m} = 1 \times 10^{-5}$. Comparing the abundances in Figure~\ref{fig-fabund_es}a and the ones from the {\it ws}-process (Figure~\ref{fig-fabund_ws}), the production of heavier nuclei beyond the Sr peak is significantly enhanced due to the effect of rotational mixing, and the overabundant region reaches up to the barium peak around $A \sim 138$ \citep{2012A&A...538L...2F}. In contrast to the {\it ws}-process, the production in the {\it es}-process strongly depends on the assumed rate for $^{17}{\rm O}(\alpha,\gamma)^{21}{\rm Ne}$. The case with reduced neutron poison efficiently produces much more intermediate mass $s$-process isotopes ($A > 90$).

Another comparison is shown in Figure~\ref{fig-fabund_es}c, which is the ratio of the final abundances between the rotating metal-poor star ({\tt z2r4}) and the even more metal-poor case ({\tt z4r4}). In both models, we adopted a reduced CF88 rate for the $\iso{O}{17}(\alpha,\gamma)\iso{Ne}{21}$ reaction (divided by $10$) in this study. The abundance ratio ({\tt z2r4}/{\tt z4r4}) is around $100$, which is the initial abundance ratio. However, in the heavier nuclei with $A>90$, the ratio is below $100$, {\it i.e.}, in the more metal-poor case ({\tt z4r4}) heavier $s$-process nuclei (beyond Sr) are more efficiently produced. In fact, focusing on the ratio of the Sr and Ba peaks, {\tt z2r4} and {\tt z4r4} show $\mbox{[Sr/Ba]}$ of $1.98$ and $1.76$, respectively. This difference is caused by the ratio of available neutrons (given by the neutron density) to the seed nuclei during the $s$-process. The lower metallicity model has effectively a higher number of neutrons compared to the initial seed abundances (mostly Fe).

The time evolution of neutron number density for the {\it es}-process ({\tt z2r4}) is shown in Figure~\ref{fig-ndens} (dashed line). Note that neutron densities are similar for the two choices of $\iso{O}{17}(\alpha,\gamma)\iso{Ne}{21}$ rates. We see that the {\it es}-process has a higher neutron density compared to the {\it ws}-process in each burning phase. Nevertheless, the nucleosynthesis flow in the {\it es}-process, shown in Figure~\ref{fig-flow_ws}, is very similar to the one in the {\it ws}-process. The dominant reaction flow consists of neutron captures and $\beta$-decays along a path following the stable isotopes. The main difference between the {\it ws}- and {\it es}-processes is that the {\it es}-process has a higher flux due to increased neutron captures and this enhances the production of heavier nuclei.

In the present study, we adopted the {\tt r2z4} model as the representative case for the following MC analysis of the {\it es}-process because its nucleosynthesis result shows the primary feature of the {\it es}-process, namely the production of the Sr and Ba peaks. The {\tt r4z4} model is also considered when discussing the uncertainty of the results due to the stellar evolution models. For both stellar models, we use a CF88 rate divided by $10$ for $\iso{O}{17}(\alpha,\gamma)\iso{Ne}{21}$ in all MC calculations.

We note that the latest evaluation in \cite{2013PhRvC..87d5805B} shows a reduction in both the $\iso{O}{17}(\alpha,\gamma)\iso{Ne}{21}$ and $\iso{O}{17}(\alpha,\gamma)\iso{Ne}{21}$ rates, but a similar $(\alpha,\mbox{n})/(\alpha,\gamma)$ reaction rate  ratio, compared to NACRE/CF88. However, these reaction rates have large uncertainties, and changes up to a factor of 10 is still reasonable. Our results show robustness to such changes so long as the {\it es}-process produces heavier $s$-process isotopes compared to the {\it ws}-process.

\subsection{The uncertainty of the {\it es}-process}
\label{sec-mc-es}

\begin{figure*}
  \centering
  \includegraphics[width=\hsize]{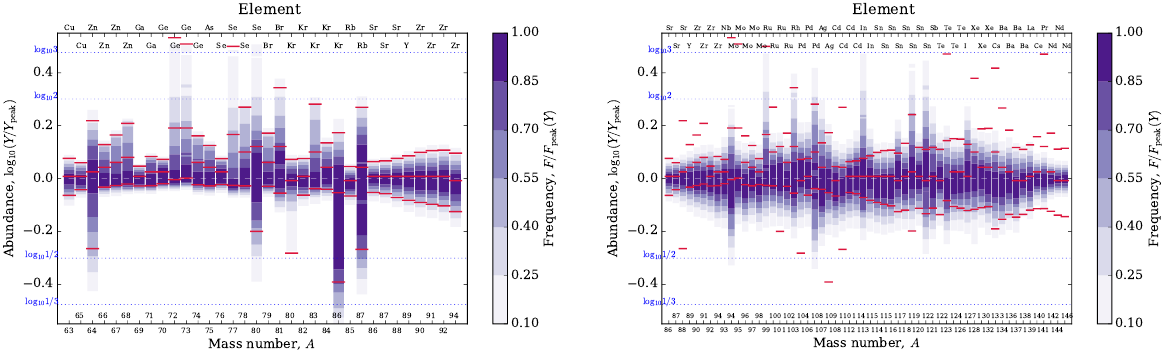}
  \caption{Uncertainty distribution in the {\it es}-process when varying all relevant neutron capture and weak rates for low mass (left) and medium mass (right) $s$-process nuclei. The colour shade is the probabilistic frequency and the 90\% probability intervals up and down marked for each nuclide (see Figure~\ref{fig-mc-hist} for examples of the distribution). Horizontal dashed lines indicate uncertainty factors of $2$, $3$, $1/2$, and $1/3$, respectively.}
   \label{fig-mc-es}
\end{figure*}

Uncertainties in {\it es}-process abundances have been determined using the same methodology as was used for the {\it ws}-process. Figure~\ref{fig-mc-es} shows the resulting production uncertainties for cases with variations of all (n,$\gamma$) reactions and weak reactions. For this plot, we choose to show stable $s$-process nuclei with $29 \leq Z \leq 40$ (left panel) and $38\leq A\leq 60$ (right panel), covering elements up to Sr and up to Ba, respectively. As in Figure~\ref{fig-mc-ws-all}, the range defined by the red lines for each isotope corresponds to $90\%$ of the abundance uncertainty distribution. The uncertainty ranges for the {\it es}-process products are also listed in Table~\ref{tab-es-unc}. Comparing Figures~\ref{fig-mc-es} and \ref{fig-mc-ws-all}, we see that the uncertainty distribution pattern is significantly different between the {\it es}-process and the {\it ws}-process, although the same nuclei exhibit a larger uncertainty in both cases. For heavier nuclei beyond Sr, the abundance uncertainty increases and is propagated from lighter to heavier nuclei.

\begin{figure*}
  \centering
  \includegraphics[width=\hsize]{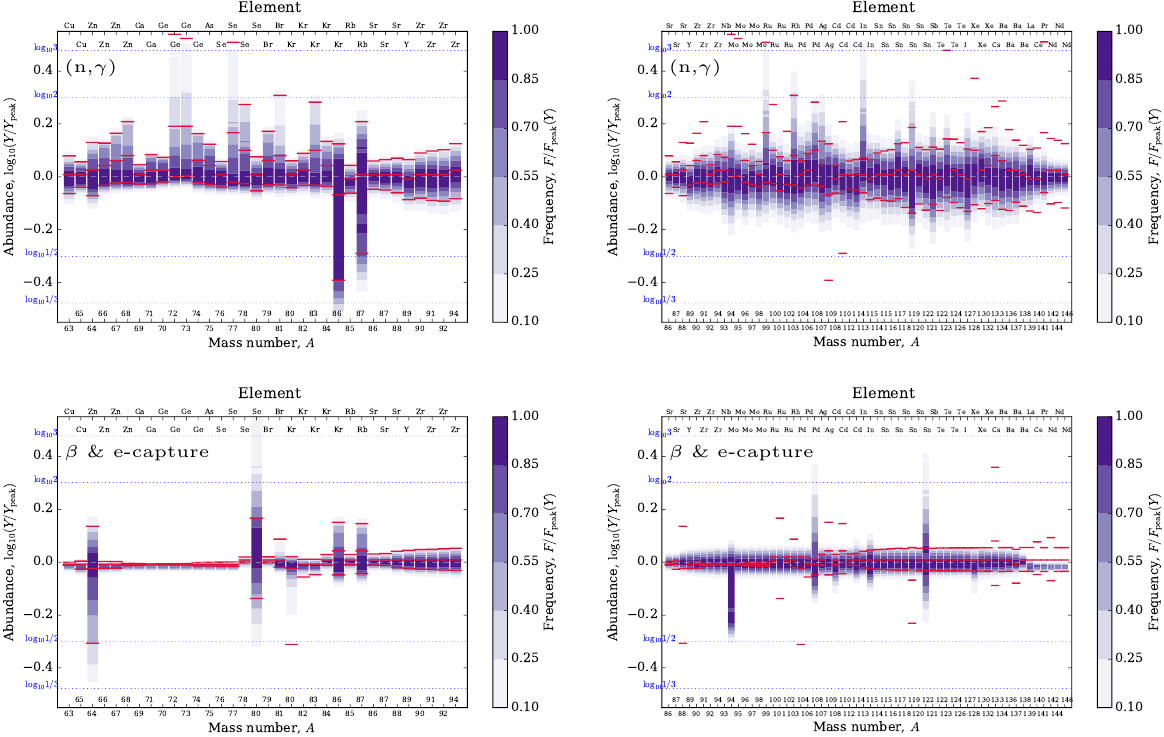}
  \caption{Same as Figure~\ref{fig-mc-es} but when varying only neutron captures (top) or weak rates (bottom).}
   \label{fig-mc-es-only}
\end{figure*}

To distinguish the individual impact on final abundance uncertainties, we also performed MC calculations with a limited number of rates being varied, {\it i.e.}, we considered (n,$\gamma$) and weak rates separately. Figure~\ref{fig-mc-es-only} shows the results obtained by only varying neutron-captures (upper panel) or weak reactions (lower panel). As already seen in the results for the {\it ws}-process, the dominant uncertainty is due to uncertainties in (n,$\gamma$), while weak rates only affect nuclei around branching points. In addition to those found in the {\it ws}-process, we identified additional such nuclei, $\iso{Nb}{94}$, $\iso{Pd}{108}$, and $\iso{Sn}{122}$, influenced by weak reactions. These species, which are intermediate mass $s$-process nuclei, are not significantly produced in the {\it ws}-process and did not appear in the results and discussion of the {\it ws}-process in Section~\ref{sec-wsproc}. The nucleosynthesis and uncertainties in the {\it es}-process are different from the ones in the {\it ws}-process and therefore we also expect different key reactions.

\begin{figure*}
  \centering
  \includegraphics[width=\hsize]{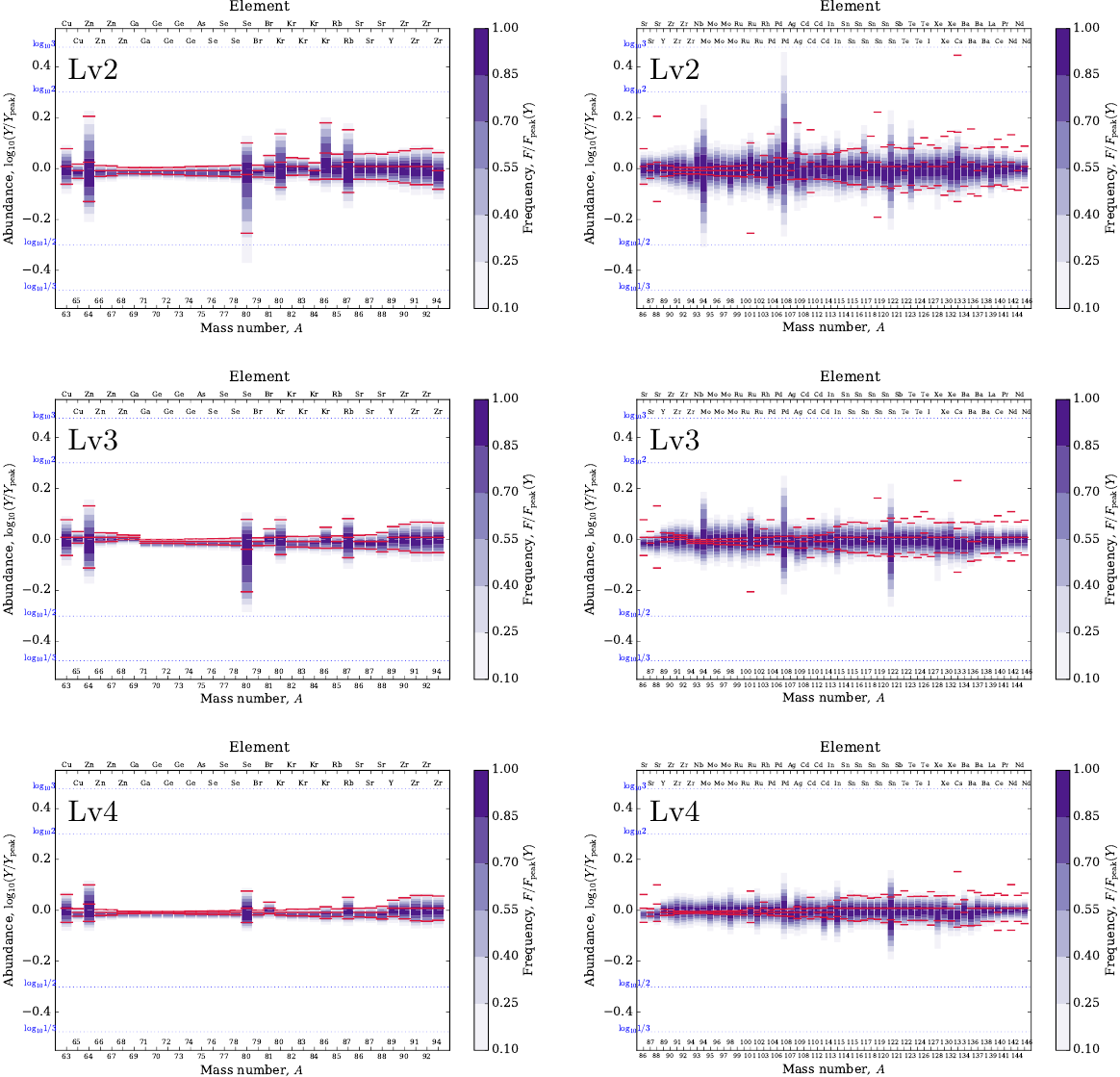}
  \caption{Results of MC calculations in the {\it es}-process for different MC levels, Lv2 (upper panel), Lv3 (middle panel), and Lv4 (lower panel). Uncertainty distributions are plotted in the same manner as in Figure~\ref{fig-mc-es}.}
   \label{fig-mc-es-lv}
\end{figure*}

\begin{table}
\centering
\caption{ \label{tab-es-unc} Uncertainties of {\it es}-process nuclei from the Lv1 MC calculation. The columns labeled ``Level'' indicates the level of the first key reaction found, as described in Section~\ref{sec-ws-key}. The remaining columns show uncertainty factors for variations Up and Down, whose values $Y_{\rm peak}(95\%)/Y_{\rm peak}$ and $Y_{\rm peak}(5\%)/Y_{\rm peak}$, respectively, enclose a 90\% probability interval, as shown in Figure~\ref{fig-mc-es}. }

\begin{tabular}{@{}lcrrlcrr}
\hline
 & Level & Up & Down & & Level & Up & Down\\
\hline
${}^{63}{\rm Cu}$ & -- &  1.19 & 0.864 & ${}^{90}{\rm Zr}$ & -- &  1.26 & 0.811 \\
${}^{65}{\rm Cu}$ &  1 &  1.15 & 0.907 & ${}^{91}{\rm Zr}$ & -- &  1.28 & 0.800 \\
${}^{64}{\rm Zn}$ &  1 &  1.65 & 0.543 & ${}^{92}{\rm Zr}$ & -- &  1.28 & 0.788 \\
${}^{66}{\rm Zn}$ &  1 &  1.34 & 0.928 & ${}^{94}{\rm Zr}$ & -- &  1.25 & 0.749 \\
${}^{67}{\rm Zn}$ &  1 &  1.46 & 0.941 & ${}^{93}{\rm Nb}$ &  2 &  1.36 & 0.760 \\
${}^{68}{\rm Zn}$ &  1 &  1.62 & 0.953 & ${}^{94}{\rm Mo}$ &  2 &  1.75 & 0.633 \\
${}^{69}{\rm Ga}$ &  1 &  1.11 & 0.937 & ${}^{95}{\rm Mo}$ & -- &  1.31 & 0.772 \\
${}^{71}{\rm Ga}$ &  1 &  1.21 & 0.940 & ${}^{96}{\rm Mo}$ &  3 &  1.29 & 0.736 \\
${}^{70}{\rm Ge}$ &  1 &  1.18 & 0.956 & ${}^{97}{\rm Mo}$ & -- &  1.31 & 0.773 \\
${}^{72}{\rm Ge}$ &  1 &  3.41 & 0.991 & ${}^{98}{\rm Mo}$ & -- &  1.29 & 0.731 \\
${}^{73}{\rm Ge}$ &  1 &  3.23 & 1.004 & ${}^{99}{\rm Ru}$ &  1 &  2.96 & 0.825 \\
${}^{74}{\rm Ge}$ &  1 &  1.45 & 0.947 & ${}^{100}{\rm Ru}$ &  3 &  1.41 & 0.784 \\
${}^{75}{\rm As}$ &  1 &  1.33 & 0.938 & ${}^{101}{\rm Ru}$ &  3 &  1.41 & 0.772 \\
${}^{76}{\rm Se}$ &  1 &  1.19 & 0.949 & ${}^{102}{\rm Ru}$ & -- &  1.35 & 0.757 \\
${}^{77}{\rm Se}$ &  1 &  3.16 & 0.942 & ${}^{103}{\rm Rh}$ &  1 &  2.39 & 0.766 \\
${}^{78}{\rm Se}$ &  1 &  1.86 & 0.938 & ${}^{104}{\rm Pd}$ &  2 &  1.54 & 0.776 \\
${}^{80}{\rm Se}$ &  1 &  4.06 & 0.631 & ${}^{106}{\rm Pd}$ &  2 &  1.55 & 0.783 \\
${}^{79}{\rm Br}$ &  1 &  1.49 & 0.935 & ${}^{108}{\rm Pd}$ &  2 &  2.62 & 0.645 \\
${}^{81}{\rm Br}$ &  1 &  2.20 & 0.881 & ${}^{107}{\rm Ag}$ &  1 &  1.85 & 0.701 \\
${}^{80}{\rm Kr}$ &  2 &  1.18 & 0.522 & ${}^{109}{\rm Ag}$ &  2 &  1.52 & 0.732 \\
${}^{82}{\rm Kr}$ &  1 &  1.19 & 0.865 & ${}^{108}{\rm Cd}$ & -- &  1.31 & 0.715 \\
${}^{83}{\rm Kr}$ &  1 &  1.91 & 0.916 & ${}^{110}{\rm Cd}$ & -- &  1.29 & 0.732 \\
${}^{84}{\rm Kr}$ &  1 &  1.36 & 0.911 & ${}^{112}{\rm Cd}$ & -- &  1.34 & 0.759 \\
${}^{86}{\rm Kr}$ &  1 &  1.49 & 0.406 & ${}^{114}{\rm Cd}$ & -- &  1.44 & 0.769 \\
${}^{85}{\rm Rb}$ &  2 &  1.11 & 0.859 & ${}^{113}{\rm In}$ &  1 &  2.96 & 0.769 \\
${}^{87}{\rm Rb}$ &  2 &  1.86 & 0.540 & ${}^{115}{\rm In}$ & -- &  1.48 & 0.752 \\
${}^{86}{\rm Sr}$ &  2 &  1.16 & 0.886 & ${}^{114}{\rm Sn}$ & -- &  1.29 & 0.727 \\
${}^{87}{\rm Sr}$ &  2 &  1.17 & 0.876 & ${}^{115}{\rm Sn}$ & -- &  1.30 & 0.718 \\
${}^{88}{\rm Sr}$ & -- &  1.19 & 0.848 & ${}^{116}{\rm Sn}$ & -- &  1.29 & 0.724 \\
${}^{89}{\rm Y}$ & -- &  1.22 & 0.828 & ${}^{117}{\rm Sn}$ &  2 &  1.58 & 0.771 \\
&&&& ${}^{118}{\rm Sn}$ & -- &  1.35 & 0.752 \\
&&&& ${}^{119}{\rm Sn}$ &  1 &  2.51 & 0.802 \\
&&&& ${}^{120}{\rm Sn}$ & -- &  1.41 & 0.776 \\
&&&& ${}^{122}{\rm Sn}$ &  2 &  2.91 & 0.715 \\
&&&& ${}^{121}{\rm Sb}$ &  1 &  1.75 & 0.723 \\
&&&& ${}^{122}{\rm Te}$ & -- &  1.41 & 0.771 \\
&&&& ${}^{123}{\rm Te}$ &  2 &  1.52 & 0.695 \\
&&&& ${}^{124}{\rm Te}$ & -- &  1.36 & 0.745 \\
&&&& ${}^{126}{\rm Te}$ & -- &  1.36 & 0.745 \\
&&&& ${}^{127}{\rm I}$ &  1 &  1.78 & 0.761 \\
&&&& ${}^{128}{\rm Xe}$ & -- &  1.46 & 0.697 \\
&&&& ${}^{130}{\rm Xe}$ & -- &  1.42 & 0.771 \\
&&&& ${}^{132}{\rm Xe}$ &  3 &  1.40 & 0.730 \\
&&&& ${}^{133}{\rm Cs}$ &  2 &  1.43 & 0.714 \\
&&&& ${}^{134}{\rm Ba}$ & -- &  1.31 & 0.718 \\
&&&& ${}^{136}{\rm Ba}$ & -- &  1.36 & 0.754 \\
&&&& ${}^{137}{\rm Ba}$ & -- &  1.30 & 0.732 \\
&&&& ${}^{138}{\rm Ba}$ & -- &  1.28 & 0.810 \\
&&&& ${}^{139}{\rm La}$ &  2 &  1.32 & 0.798 \\
&&&& ${}^{140}{\rm Ce}$ & -- &  1.21 & 0.867 \\
&&&& ${}^{141}{\rm Pr}$ &  2 &  1.22 & 0.851 \\
&&&& ${}^{142}{\rm Nd}$ & -- &  1.16 & 0.899 \\
&&&& ${}^{144}{\rm Nd}$ &  3 &  1.15 & 0.911 \\
&&&& ${}^{146}{\rm Nd}$ & -- &  1.14 & 0.912 \\
\hline
\end{tabular}
\medskip
\end{table}

\begin{table*}
\centering
\caption{ \label{tab-es-key}The key reaction rates for the {\it es}-process. Similar to Table~\ref{tab-ws-key}, key rates in levels 1--3 are shown, along with their correlation factors $r_{\rm cor,0}$, $r_{\rm cor,1}$, and $r_{\rm cor,2}$, respectively. Significant correlation values are underlined. Also shown for each rate are the g.s.\ contributions to the stellar rate for neutron captures and uncertainty factors of the $\beta$-decay rates at two plasma temperatures, respectively.}
\begin{tabular}{@{}l|c|ccccccccc}
\hline
Nuclide & $r_{{\rm cor},0}$ & $r_{{\rm cor},1}$ & $r_{{\rm cor},2}$& Key Rate & Key Rate & Key Rate& $X_0$ & Weak Rate \\
&&&& Level 1 & Level 2 & Level 3 & $(8$, $30~{\rm keV})$ & $(8$, $30~{\rm keV})$\\
\hline
$\iso{Cu}{65}$ & \underline{-0.83} &       &       & $\iso{Cu}{65}(\mbox{n},\gamma)\iso{Cu}{66}$& $ $& $ $&1.00, 1.00&\\
$\iso{Zn}{64}$ & \underline{ 0.72} &       &       & $\iso{Cu}{64}(\beta^-)\iso{Zn}{64}$& $ $& $ $& & 1.30, 1.36\\
     & -0.45 & \underline{-0.67} &       & $ $& $\iso{Cu}{64}({\rm e}^-,\nu_{\rm e})\iso{Ni}{64}$& $ $& & ${\rm e}^-$ capture\\
     & -0.36 & -0.52 & \underline{-0.72} & $ $& $ $& $\iso{Zn}{64}(\mbox{n},\gamma)\iso{Zn}{65}$&1.00, 1.00&\\
$\iso{Zn}{66}$ & \underline{-0.96} &       &       & $\iso{Zn}{66}(\mbox{n},\gamma)\iso{Zn}{67}$& $ $& $ $&1.00, 1.00&\\
     & -0.13 & -0.58 & \underline{-0.67} & $ $& $ $& $\iso{Fe}{57}(\mbox{n},\gamma)\iso{Fe}{58}$&0.73, 0.59&\\
$\iso{Zn}{67}$ & \underline{-0.97} &       &       & $\iso{Zn}{67}(\mbox{n},\gamma)\iso{Zn}{68}$& $ $& $ $&1.00, 1.00&\\
$\iso{Zn}{68}$ & \underline{-0.98} &       &       & $\iso{Zn}{68}(\mbox{n},\gamma)\iso{Zn}{69}$& $ $& $ $&1.00, 1.00&\\
$\iso{Ga}{69}$ & \underline{-0.92} &       &       & $\iso{Ga}{69}(\mbox{n},\gamma)\iso{Ga}{70}$& $ $& $ $&1.00, 1.00&\\
$\iso{Ga}{71}$ & \underline{-0.97} &       &       & $\iso{Ga}{71}(\mbox{n},\gamma)\iso{Ga}{72}$& $ $& $ $&1.00, 1.00&\\
$\iso{Ge}{70}$ & \underline{-0.95} &       &       & $\iso{Ge}{70}(\mbox{n},\gamma)\iso{Ge}{71}$& $ $& $ $&1.00, 1.00&\\
$\iso{Ge}{72}$ & \underline{-0.94} &       &       & $\iso{Ge}{72}(\mbox{n},\gamma)\iso{Ge}{73}$& $ $& $ $&1.00, 1.00&\\
$\iso{Ge}{73}$ & \underline{-0.94} &       &       & $\iso{Ge}{73}(\mbox{n},\gamma)\iso{Ge}{74}$& $ $& $ $&0.88, 0.81&\\
     &  0.03 & \underline{ 0.82} &       & $ $& $\iso{Ni}{64}(\mbox{n},\gamma)\iso{Ni}{65}$& $ $&1.00, 1.00&\\
$\iso{Ge}{74}$ & \underline{-0.97} &       &       & $\iso{Ge}{74}(\mbox{n},\gamma)\iso{Ge}{75}$& $ $& $ $&1.00, 1.00&\\
$\iso{As}{75}$ & \underline{-0.96} &       &       & $\iso{As}{75}(\mbox{n},\gamma)\iso{As}{76}$& $ $& $ $&1.00, 1.00&\\
$\iso{Se}{76}$ & \underline{-0.90} &       &       & $\iso{Se}{76}(\mbox{n},\gamma)\iso{Se}{77}$& $ $& $ $&1.00, 1.00&\\
$\iso{Se}{77}$ & \underline{-0.93} &       &       & $\iso{Se}{77}(\mbox{n},\gamma)\iso{Se}{78}$& $ $& $ $&1.00, 1.00&\\
$\iso{Se}{78}$ & \underline{-0.97} &       &       & $\iso{Se}{78}(\mbox{n},\gamma)\iso{Se}{79}$& $ $& $ $&1.00, 1.00&\\
     &  0.07 &  0.46 & \underline{ 0.70} & $ $& $ $& $\iso{Fe}{56}(\mbox{n},\gamma)\iso{Fe}{57}$&1.00, 1.00&\\
$\iso{Se}{80}$ & \underline{-0.78} &       &       & $\iso{Br}{80}(\beta^-)\iso{Kr}{80}$& $ $& $ $& & 1.31, 4.70\\
     &  0.18 &  0.47 & \underline{ 0.89} & $ $& $ $& $\iso{Br}{80}({\rm e}^-,\nu_{\rm e})\iso{Se}{80}$& & ${\rm e}^-$ capture\\
$\iso{Br}{79}$ & \underline{-0.96} &       &       & $\iso{Br}{79}(\mbox{n},\gamma)\iso{Br}{80}$& $ $& $ $&1.00, 1.00&\\
$\iso{Br}{81}$ & \underline{-0.86} &       &       & $\iso{Kr}{81}(\mbox{n},\gamma)\iso{Kr}{82}$& $ $& $ $&1.00, 0.98&\\
$\iso{Kr}{80}$ & -0.28 & \underline{-0.78} &       & $ $& $\iso{Br}{80}(\beta^+)\iso{Se}{80}$& $ $\\
     & -0.30 & -0.43 & \underline{-0.67} & $ $& $ $& $\iso{Kr}{80}(\mbox{n},\gamma)\iso{Kr}{81}$&1.00, 1.00&\\
$\iso{Kr}{82}$ & \underline{-0.78} &       &       & $\iso{Kr}{82}(\mbox{n},\gamma)\iso{Kr}{83}$& $ $& $ $&1.00, 1.00&\\
$\iso{Kr}{83}$ & \underline{-0.95} &       &       & $\iso{Kr}{83}(\mbox{n},\gamma)\iso{Kr}{84}$& $ $& $ $&0.81, 0.74&\\
$\iso{Kr}{84}$ & \underline{-0.88} &       &       & $\iso{Kr}{84}(\mbox{n},\gamma)\iso{Kr}{85}$& $ $& $ $&1.00, 1.00&\\
$\iso{Kr}{86}$ & \underline{ 0.87} &       &       & $\iso{Kr}{85}(\mbox{n},\gamma)\iso{Kr}{86}$& $ $& $ $&1.00, 1.00&\\
$\iso{Rb}{85}$ & -0.62 & \underline{-0.73} &       & $ $& $\iso{Rb}{85}(\mbox{n},\gamma)\iso{Rb}{86}$& $ $&1.00, 1.00&\\
$\iso{Rb}{87}$ & -0.35 & \underline{-0.74} &       & $ $& $\iso{Kr}{85}(\beta^-)\iso{Rb}{85}$& $ $& & 1.30, 1.30\\
     &  0.22 &  0.44 & \underline{ 0.75} & $ $& $ $& $\iso{Kr}{86}(\mbox{n},\gamma)\iso{Kr}{87}$&1.00, 1.00&\\
$\iso{Sr}{86}$ & -0.57 & \underline{-0.67} &       & $ $& $\iso{Sr}{86}(\mbox{n},\gamma)\iso{Sr}{87}$& $ $&1.00, 1.00&\\
$\iso{Sr}{87}$ & -0.55 & \underline{-0.66} &       & $ $& $\iso{Sr}{87}(\mbox{n},\gamma)\iso{Sr}{88}$& $ $&1.00, 1.00&\\
$\iso{Nb}{93}$ & -0.59 & \underline{-0.76} &       & $ $& $\iso{Zr}{93}(\mbox{n},\gamma)\iso{Zr}{94}$& $ $&1.00, 1.00&\\
$\iso{Mo}{94}$ &  0.64 & \underline{ 0.68} &       & $ $& $\iso{Zr}{93}(\beta^-)\iso{Nb}{93}$& $ $& & 1.30, 1.30\\
     & -0.47 & -0.51 & \underline{-0.88} & $ $& $ $& $\iso{Mo}{94}(\mbox{n},\gamma)\iso{Mo}{95}$&1.00, 1.00&\\
$\iso{Mo}{96}$ & -0.42 & -0.58 & \underline{-0.66} & $ $& $ $& $\iso{Mo}{96}(\mbox{n},\gamma)\iso{Mo}{97}$&1.00, 1.00&\\
$\iso{Ru}{99}$ & \underline{-0.86} &       &       & $\iso{Ru}{99}(\mbox{n},\gamma)\iso{Ru}{100}$& $ $& $ $&1.00, 1.00&\\
$\iso{Ru}{100}$ & -0.44 & -0.61 & \underline{-0.69} & $ $& $ $& $\iso{Ru}{100}(\mbox{n},\gamma)\iso{Ru}{101}$&1.00, 1.00&\\
$\iso{Ru}{101}$ & -0.47 & -0.65 & \underline{-0.73} & $ $& $ $& $\iso{Ru}{101}(\mbox{n},\gamma)\iso{Ru}{102}$&1.00, 1.00&\\
$\iso{Rh}{103}$ & \underline{-0.85} &       &       & $\iso{Rh}{103}(\mbox{n},\gamma)\iso{Rh}{104}$& $ $& $ $&0.95, 0.80&\\
$\iso{Pd}{104}$ & -0.60 & \underline{-0.77} &       & $ $& $\iso{Pd}{104}(\mbox{n},\gamma)\iso{Pd}{105}$& $ $&1.00, 1.00&\\
$\iso{Pd}{106}$ & -0.60 & \underline{-0.78} &       & $ $& $\iso{Pd}{106}(\mbox{n},\gamma)\iso{Pd}{107}$& $ $&1.00, 1.00&\\
$\iso{Pd}{108}$ & -0.61 & \underline{-0.66} &       & $ $& $\iso{Pd}{107}(\beta^-)\iso{Ag}{107}$& $ $& & 1.30, 1.36\\
     & -0.47 & -0.50 & \underline{-0.75} & $ $& $ $& $\iso{Pd}{108}(\mbox{n},\gamma)\iso{Pd}{109}$&1.00, 1.00&\\
$\iso{Ag}{107}$ & \underline{-0.80} &       &       & $\iso{Ag}{107}(\mbox{n},\gamma)\iso{Ag}{108}$& $ $& $ $&1.00, 1.00&\\
$\iso{Ag}{109}$ & -0.56 & \underline{-0.71} &       & $ $& $\iso{Ag}{109}(\mbox{n},\gamma)\iso{Ag}{110}$& $ $&1.00, 1.00&\\
$\iso{In}{113}$ & \underline{-0.85} &       &       & $\iso{In}{113}(\mbox{n},\gamma)\iso{In}{114}$& $ $& $ $&1.00, 1.00&\\
$\iso{Sn}{117}$ & -0.58 & \underline{-0.77} &       & $ $& $\iso{Sn}{117}(\mbox{n},\gamma)\iso{Sn}{118}$& $ $&1.00, 1.00&\\
$\iso{Sn}{119}$ & \underline{-0.83} &       &       & $\iso{Sn}{119}(\mbox{n},\gamma)\iso{Sn}{120}$& $ $& $ $&0.89, 0.75&\\
$\iso{Sn}{122}$ & \underline{-0.68} &       &       & $\iso{Sb}{122}(\beta^-)\iso{Te}{122}$& $ $& $ $& & 1.30, 2.81\\
     & -0.32 & -0.64 & \underline{-0.67} & $ $& $ $& $\iso{Sb}{122}(\beta^-)\iso{Te}{122}$& & 1.30, 2.81\\
$\iso{Sb}{121}$ & \underline{-0.73} &       &       & $\iso{Sb}{121}(\mbox{n},\gamma)\iso{Sb}{122}$& $ $& $ $&0.98, 0.93&\\
$\iso{Te}{123}$ & -0.64 & \underline{-0.83} &       & $ $& $\iso{Te}{123}(\mbox{n},\gamma)\iso{Te}{124}$& $ $&1.00, 1.00&\\
$\iso{I}{127}$ & \underline{-0.70} &       &       & $\iso{I}{127}(\mbox{n},\gamma)\iso{I}{128}$& $ $& $ $&1.00, 0.99&\\
$\iso{Xe}{132}$ & -0.37 & -0.58 & \underline{-0.66} & $ $& $ $& $\iso{Xe}{132}(\mbox{n},\gamma)\iso{Xe}{133}$&1.00, 1.00&\\
$\iso{Cs}{133}$ & -0.49 & \underline{-0.70} &       & $ $& $\iso{Cs}{133}(\mbox{n},\gamma)\iso{Cs}{134}$& $ $&1.00, 1.00&\\
$\iso{La}{139}$ & -0.56 & \underline{-0.73} &       & $ $& $\iso{La}{139}(\mbox{n},\gamma)\iso{La}{140}$& $ $&1.00, 1.00&\\
$\iso{Pr}{141}$ & -0.56 & \underline{-0.66} &       & $ $& $\iso{Pr}{141}(\mbox{n},\gamma)\iso{Pr}{142}$& $ $&1.00, 1.00&\\
$\iso{Nd}{144}$ &  0.51 &  0.61 & \underline{ 0.65} & $ $& $ $& $\iso{Ba}{138}(\mbox{n},\gamma)\iso{Ba}{139}$&1.00, 1.00&\\
\hline
\end{tabular}
\end{table*}

\begin{table}
\centering
\caption{ \label{tab-es-unc-z} Uncertainties in the {\it es}-process for elemental abundances in the {\tt z2r4} model. The columns `Up' and `Down' correspond to the upper and lower boundary of the uncertainty range similar to Table~\ref{tab-es-unc} but for each element. The correlation coefficient $r_{\rm cor}$ and the corresponding reaction is shown for Lv1 key reactions (See Section~\ref{sec-es-key}).}

\begin{tabular}{@{}ccccc}
\hline
Element & Up & Down & $r_{\rm cor}$ & Key reaction \\
\hline
  Cu &  1.14 & 0.913 & -0.73 & ${}^{65}{\rm{Cu}} (\mbox{n},\gamma) ^{66}{\rm{Cu}}$ \\
  Zn &  1.28 & 0.900 & -0.91 & ${}^{68}{\rm{Zn}} (\mbox{n},\gamma) {}^{69}{\rm{Zn}}$ \\
  Ga &  1.11 & 0.935 & -0.83 & ${}^{71}{\rm{Ga}} (\mbox{n},\gamma) {}^{72}{\rm{Ga}}$ \\
  Ge &  1.28 & 0.852 & -0.74 & ${}^{72}{\rm{Ge}} (\mbox{n},\gamma) {}^{73}{\rm{Ge}}$ \\
  As &  1.33 & 0.938 & -0.96 & ${}^{75}{\rm{As}} (\mbox{n},\gamma) {}^{76}{\rm{As}}$ \\
  Se &  1.41 & 0.828 & -0.73 & ${}^{78}{\rm{Se}} (\mbox{n},\gamma) {}^{79}{\rm{Se}}$ \\
  Br &  1.51 & 0.851 & -0.80 & ${}^{81}{\rm{Kr}} (\mbox{n},\gamma) {}^{82}{\rm{Kr}}$ \\
  Kr &  1.19 & 0.869 &        & $                              $ \\
  Rb &  1.19 & 0.867 &        & $                              $ \\
  Sr &  1.18 & 0.861 &        & $                              $ \\
   Y &  1.22 & 0.828 &        & $                              $ \\
  Zr &  1.25 & 0.808 &        & $                              $ \\
  Nb &  1.36 & 0.760 &        & $                              $ \\
  Mo &  1.26 & 0.747 &        & $                              $ \\
  Ru &  1.39 & 0.793 &        & $                              $ \\
  Rh &  2.39 & 0.766 & -0.85 & ${}^{103}{\rm{Rh}} (\mbox{n},\gamma) {}^{104}{\rm{Rh}}$ \\
  Pd &  1.36 & 0.744 &        & $                              $ \\
  Ag &  1.36 & 0.686 &        & $                              $ \\
  Cd &  1.34 & 0.761 &        & $                              $ \\
  In &  2.75 & 0.743 & -0.85 & ${}^{113}{\rm{In}} (\mbox{n},\gamma) {}^{114}{\rm{In}}$ \\
  Sn &  1.35 & 0.753 &        & $                              $ \\
  Sb &  1.75 & 0.723 & -0.73 & ${}^{121}{\rm{Sb}} (\mbox{n},\gamma) {}^{122}{\rm{Sb}}$ \\
  Te &  1.42 & 0.769 &        & $                              $ \\
   I &  1.78 & 0.761 & -0.70 & ${}^{127}{\rm{I}} (\mbox{n},\gamma) {}^{128}{\rm{I}}$ \\
  Xe &  1.43 & 0.767 &        & $                              $ \\
  Cs &  1.43 & 0.714 &        & $                              $ \\
  Ba &  1.31 & 0.785 &        & $                              $ \\
  La &  1.32 & 0.798 &        & $                              $ \\
  Ce &  1.21 & 0.867 &        & $                              $ \\
  Pr &  1.22 & 0.851 &        & $                              $ \\
  Nd &  1.15 & 0.907 &        & $                              $ \\
\hline
\end{tabular}
\medskip
\end{table}

\begin{table}
\centering
\caption{\label{tab-es-unc-z-lz} Uncertainty and key reactions of {\it es}-process elements ({\tt z4r4}). The columns are the same as Table~\ref{tab-es-unc-z}.}
\begin{tabular}{@{}ccccc}
\hline
Element & Up & Down & $r_{\rm cor}$ & Key reaction \\
\hline
  Cu &  1.15 & 0.893 & -0.66 & $^{65}$Cu$(\mbox{n},\gamma)^{66}$Cu \\
  Zn &  1.30 & 0.892 & -0.90 & $^{68}$Zn$(\mbox{n},\gamma)^{69}$Zn \\
  Ga &  1.12 & 0.930 & -0.79 & $^{71}$Ga$(\mbox{n},\gamma)^{72}$Ga \\
  Ge &  1.29 & 0.849 & -0.74 & $^{72}$Ge$(\mbox{n},\gamma)^{73}$Ge \\
  As &  1.33 & 0.940 & -0.97 & $^{75}$As$(\mbox{n},\gamma)^{76}$As \\
  Se &  1.53 & 0.901 & -0.74 & $^{78}$Se$(\mbox{n},\gamma)^{79}$Se \\
  Br &  1.45 & 0.823 & -0.80 & $^{81}$Kr$(\mbox{n},\gamma)^{82}$Kr \\
  Kr &  1.18 & 0.871 &        &                                \\
  Rb &  1.19 & 0.874 &        &                                \\
  Sr &  1.16 & 0.879 &        &                                \\
   Y &  1.21 & 0.841 &        &                                \\
  Zr &  1.20 & 0.785 &        &                                \\
  Nb &  1.36 & 0.760 &        &                                \\
  Mo &  1.31 & 0.778 &        &                                \\
  Ru &  1.39 & 0.791 &        &                                \\
  Rh &  2.39 & 0.756 & -0.85 & $^{103}$Rh$~(\mbox{n},\gamma)~$$^{104}$Rh \\
  Pd &  1.37 & 0.748 &        &                                \\
  Ag &  1.49 & 0.744 &        &                                \\
  Cd &  1.35 & 0.757 &        &                                \\
  In &  2.83 & 0.776 & -0.85 & $^{113}$In$~(\mbox{n},\gamma)~$$^{114}$In \\
  Sn &  1.36 & 0.751 &        &                                \\
  Sb &  1.82 & 0.738 & -0.71 & $^{121}$Sb$~(\mbox{n},\gamma)~$$^{122}$Sb \\
  Te &  1.40 & 0.731 &        &                                \\
   I &  1.74 & 0.721 & -0.69 & $^{127}$I$~(\mbox{n},\gamma)~$$^{128}$I \\
  Xe &  1.46 & 0.754 &        &                                \\
  Cs &  1.52 & 0.730 &        &                                \\
  Ba &  1.43 & 0.781 &        &                                \\
  La &  1.49 & 0.820 &        &                                \\
  Ce &  1.34 & 0.853 &        &                                \\
  Pr &  1.30 & 0.815 &        &                                \\
  Nd &  1.22 & 0.864 &        &                                \\
\hline
\end{tabular}
\medskip
\end{table}

\subsection{Key reactions in the {\it es}-process}
\label{sec-es-key}

As for the {\it ws}-process, we identified key reactions with a strong influence on the final abundance uncertainties. The key reactions for the {\it es}-process with a high correlation ($|r_{\rm cor}| \ge 0.65$) are listed in Table~\ref{tab-es-key}. The list includes $30$ Lv1 key reactions in the {\it es}-process. A majority of key reactions are neutron captures along the $s$-process path, while only a few weak reactions around branchings have an impact: only $\beta^{-}$-decay of $\iso{Cu}{64}$, $\iso{Br}{80}$, and $\iso{Sb}{122}$ are listed at Lv1. Additional reactions are found at Lv2 and Lv3, based on Lv2 and Lv3 MC runs, respectively. The resulting uncertainty distributions of {\it es}-process abundances at different levels are shown in Figure~\ref{fig-mc-es-lv}. With decreasing number of varied reactions in the MC variations in the various levels, the uncertainties in the final abundances shrink and finally the Lv4 results show only small uncertainties.

The list of key {\it es}-process reactions found here is more extensive than the list of key {\it ws}-process reactions presented in Table~\ref{tab-ws-key}. While the {\it es}-process obviously has additional key reactions for intermediate mass $s$-process nuclei from the Sr peak to lighter lanthanides beyond Ba, which the {\it ws}-process does not produce, we find that the {\it es}-process has more key reactions even for the lighter $s$-process nuclei. It is common (although not in all the cases) that a reaction in the {\it es}-process has a larger correlation $|r_{\rm cor}|$ as the same reaction in the {\it ws}-process. This increase in the correlation factor is caused by the stronger reaction flow in the {\it es}-process, which enhances the uncertainty propagation.

Observationally, the isotopic composition of neutron capture elements in galactic stars has been measured only for few elements ({\it e.g.,} Ba, Sm, Nd and Eu) so far \citep{2008AIPC..990..172R, 2015A&A...579A..94G}, whereas the elemental abundances are available for much more elements (with the notable exception of In). The uncertainty ranges and primary key reactions for each {\it es}-process element are summarized in Table~\ref{tab-es-unc-z}. While most elements have an uncertainty range up to a factor of $1.5$, Rh and In show significantly higher uncertainties, exceeding a factor of two. Interestingly, both Rh and In have a key neutron-capture reaction. The element Rh has only one stable isotope, $\iso{Rh}{103}$, so that the key reaction for the elemental production obviously corresponds to the one for this isotope as given in Table~\ref{tab-es-key}. In very high quality spectra of galactic halo stars, the abundance of rhodium can be measured \citep{2002A&A...387..560H} and can be used to constrain the key reaction rate, $\iso{Rh}{103}(\mbox{n},\gamma)\iso{Rh}{104}$. For indium, the $\iso{In}{113}$ isotope is mostly produced by the $s$-process\footnote{In the solar abundances, the isotopic contribution of $\iso{In}{113}$ to the elemental abundance is less than 5\% and  $\iso{In}{115}$ is the dominant isotope. This isotopic comparison is driven by another nucleosynthesis process, namely the $r$-process, that produces most of the solar In.}, where the mass fraction of $\iso{In}{113}$ is 0.964 of the totally produced In. Thus, the key reaction of $\iso{In}{113}$ naturally has a significant impact on uncertainty of In.

In addition to the {\it es}-process in our standard model ({\tt z2r4}), the uncertainty range and the key reactions for the very low metal-poor star, {\tt z4r4}, are shown in Table~\ref{tab-es-unc-z-lz}. This model showed a different final abundance distribution (Figure~\ref{fig-fabund_es}). Nevertheless, comparing Table~\ref{tab-es-unc-z} and Table~\ref{tab-es-unc-z-lz}, we do not find any significant differences either in the uncertainty ranges or in correlation coefficients and listed key reactions. This indicates that our conclusions regarding the nuclear physics uncertainties in the {\it es}-process are robust with only a weak dependence on the stellar models as long as the final abundance distribution shows a typical {\it es}-process pattern, as discussed in Section~\ref{sec-es-basis}.

\section{Opportunities for improved nuclear data}
\label{sec-nuc-data}

Tables~\ref{tab-ws-key} and~\ref{tab-es-key} list the key nuclear reaction rates identified in this study for the {\it ws}- and {\it es}-processes, respectively. Uncertainties in these rates have the greatest overall impact on final abundances, and are therefore prioritised for future precision measurement. Although neutron captures on stable or long-lived nuclei can, in principle, be measured, it is not always possible to also experimentally constrain the \textit{stellar} neutron capture rates which contain contributions from reactions on thermally excited states of the target nucleus. As highlighted earlier, excited state contributions are important for some nuclei even at $s$-process temperatures. This should be kept in mind when selecting reactions from Tables~\ref{tab-ws-key} and~\ref{tab-es-key} for future experiments. To simplify the task, the g.s.\ contributions to the stellar rates are also given in the tables of the key reactions. The larger the ground-state (g.s.) contribution, the better an experiment can constrain a stellar rate, as can also be seen from Equation~(\ref{eq-unc_ng}). Since many key reactions have a ground-state contribution close to one, there is a good prospect of future experiments reducing the uncertainties in these rates.

The experimental measurement of (n,\,$\gamma$) type reactions is well established, for example through the use of activation or neutron time of flight techniques. For direct measurements, one requires a radioactively stable or long-lived target of sufficient mass and isotopic purity, and in addition a solid and chemically inert target is preferred. Compounds may be used to satisfy the latter requirement. Many of the reactions listed in tables~\ref{tab-ws-key} and~\ref{tab-es-key} satisfy these requirements, and indeed many have been measured, although with greatly varying levels of completeness, precision, and consistency. We used the KADoNiS database \citep{2006AIPC..819..123D} to define the standard neutron capture rates for our MC variations. Opportunities for improvements to the library of nuclear data are extensive. For example, the $^{77}$Se and $^{78}$Se(n,\,$\gamma$) reactions are identified as Lv1 key reactions in both the {\it ws}- and {\it es}-processes. In the case of $^{77}$Se, existing precision data cover only the $15 < E < 100$~keV and around the 510~keV regions. Recent work \citep{kamada2010measurements} revealed inconsistencies at the level of 10--20\% as compared to earlier evaluations. Similarly, for $\iso{Se}{78}({\mbox n}, \gamma)\iso{Se}{79}$, a previous activation study \citep{2006isna.confE..89D} and time of flight measurements \citep[][and private communication]{igashira2011} show a large discrepancy in values of Maxwellian averaged cross section.

For some of the reactions identified here there are presently no experimental data available, {\it e.g.}, neutron captures on $^{80}$Br, $^{81}$Kr, $^{99}$Ru, and $^{103}$Ru, although $^{80}$Br and $^{103}$Ru are radioactive isotopes with half lives of 17.68 m and 39.25 d, respectively.


\section{Summary and Conclusions}
\label{sec-summary}

We investigated the impact of nuclear-physics uncertainties on the $s$-process in massive stars, focusing on neutron captures and weak reaction (mostly $\beta$-decays) rates. Adopting the evolution models of a solar metallicity star and a fast rotating metal-poor star, we studied the {\it ws}-process and rotation-induced {\it es}-process, respectively. Using newly evaluated temperature-dependent uncertainties for neutron capture and $\beta$-decay rates, we performed a series of MC calculations with a nuclear reaction network. We obtained complete information about the uncertainty of final abundances and the identity of the underlying key reaction rates.
The results are summarised as follows.
\begin{enumerate}

\item For both of the {\it ws}- and {\it es}-processes, the uncertainty range in the final abundances (the upper and lower boundaries of 90\% probability around the mean value) is relatively small for the majority of $s$-process nuclei, typically within a few tens of percent. Several nuclei have a larger uncertainty in the final abundance, which is beyond a factor of $2$ but is less than a factor of $5$. In general, the resulting frequency distribution of the final abundances is continuous but asymmetric in shape.

\item Our MC calculations have determined the correlation ($r_{\rm cor}$ defined by Equation~\ref{eq-cor}) between the reaction rate variation and the final abundances, identifying key neutron capture reactions and $\beta$-decays ($|r_{\rm cor}| \ge 0.65$). We find $10$ important reactions to improve the {\it ws}-process abundance prediction as well as $30$ rates for the {\it es}-process. In addition, there are $11$ and $32$ extra rates of secondary importance for the {\it ws}- and {\it es}-processes, respectively, summarised in Table~\ref{tab-ws-key} and \ref{tab-es-key}.

\item The {\it es}-process exhibits different features of uncertainty distribution and key reactions from the {\it ws}-process. Even within the lighter $s$-process nuclei region, overlapped with {\it ws}-process products, the {\it es}-process has a larger uncertainty with more key reactions. This is caused by a stronger flow of nucleosynthesis due to a higher neutron exposure.

\item Although there are still uncertainties in stellar models for the {\it es}-process, the results of MC variation based on models at two different metallicities lead to the same key reaction rate list. Thus the key reaction rate list obtained in this study is robust, as long as the final abundances show a ``typical'' es-process pattern, producing intermediate s-process nuclei from strontium to barium.

\end{enumerate}

In the priority list (key reactions) for both {\it ws}- and {\it es}-processes, there are some reactions for which future experiments can reduce the current uncertainty. The weak reaction rates are mostly from theory and their uncertainty is significant, especially at the stellar temperature. Improved relevant nuclear physics properties and theoretical predictions are desirable. Conversely, our MC results in combination with future observations may be able to provide constraints on the nuclear physics.

For the {\it es}-process in particular there are only limited observational constraints. However, we find that the uncertainty range and the key reactions are similar in a certain range of metallicities for fast rotating evolution models. The evaluated uncertainty in each element can be immediately useful for some astronomical application, {\it e.g.} comparison to abundance observations in metal-poor stars and  theoretical calculations of galactic chemical evolution. We expect that such astronomical comparisons will provide further restriction to the {\it es}-process abundances and relevant nuclear reactions.

As demonstrated in this study of the impact of (n,$\gamma$) rate uncertainties on the $s$-process production in massive stars, a Monte Carlo framework provides a robust tool for the analysis of uncertainties. This is the second paper in a series of applications of our recently developed MC framework, including sets of realistic variation limits, which was described in more detail in the first paper \citep{2016MNRAS.463.4153R}. The framework is applicable to further nucleosynthesis processes, in particular to the synthesis of heavy elements in which a large number of reactions are involved. A study of the main $s$-process is underway (G.~Cescutti~et~al, in preparation). Applications of the MC framework to further nucleosynthesis processes, such as the $\gamma$-process in thermonuclear supernovae, the $r$- and the $\nu${\it p}-processes are planned.

\section*{Acknowledgments}

The authors thank U.~Frischknecht and C.~Winteler for contributing to the development of the MC code. This project has been financially supported by European Research Council (EU-FP7-ERC-2012-St Grant 306901-SHYNE and GA 321263-FISH) and Swiss National Science Foundation. Numerical computations were in part carried out on the Cambridge COSMOS SMP system (part of the STFC DiRAC HPC Facility supported by BIS NeI capital grant ST/J005673/1 and STFC grants ST/H008586/1, ST/K00333X/1) and Edinburgh Compute and Data Facility, Eddie mark 3, which has support from the eDIKT initiative. N.N. used computer facilities at CfCA, National Astronomical Observatory of Japan. R.H. acknowledges support from the World Premier International Research Center Initiative, MEXT, Japan. G.C. acknowledges financial support from the European Union Horizon 2020 research and innovation programme under the Marie Sk{\l}odowska-Curie grant agreement No 664931. The University of Edinburgh is a charitable body, registered in Scotland, with registration No. SC005336.



\bibliographystyle{mnras}
\bibliography{ref}

\bsp	
\label{lastpage}
\end{document}